%% file: main.tex
\documentclass[sigconf]{acmart}
\AtBeginDocument{%
  \providecommand\BibTeX{{%
    \normalfont B\kern-0.5em{\scshape i\kern-0.25em b}\kern-0.8em\TeX}}}

\copyrightyear{2025}
\acmYear{2025}
\setcopyright{cc}
\setcctype{by}
\acmConference[CHI '25]{CHI Conference on Human Factors in Computing
Systems}{April 26-May 1, 2025}{Yokohama, Japan}
\acmBooktitle{CHI Conference on Human Factors in Computing Systems (CHI
'25), April 26-May 1, 2025, Yokohama,
Japan}\acmDOI{10.1145/3706598.3713776}
\acmISBN{979-8-4007-1394-1/25/04}

\usepackage{color, colortbl}
\usepackage{subfig}
\usepackage{multirow}
\usepackage{longtable}
\usepackage[font=small,labelfont=bf]{caption}
\usepackage{wasysym}
\usepackage[nolist,nohyperlinks, printonlyused, withpage]{acronym}
\usepackage{enumitem}

\begin{document}


\title[How People With and Without ADHD Recognise and Avoid Dark Patterns on Social Media]{A Comparative Study of How People With and Without ADHD Recognise and Avoid Dark Patterns on Social Media}



%

\author{Thomas Mildner}
\orcid{0000-0002-1712-0741}
\affiliation{%
  \institution{University of Bremen}
  \city{Bremen}
  \country{Germany}
  \postcode{28215}
}
\email{mildner@uni-bremen.de}

\author{Daniel Fidel}
\orcid{0009-0007-4109-5826}
\affiliation{
    \institution{University of Bremen}
    \city{Bremen}
    \country{Germany}
}
\email{daniel.fidel.df@proton.me}

\author{Evropi Stefanidi}
\orcid{0000-0001-5547-6426}
\affiliation{%
  \institution{University of Bremen}
  \city{Bremen}
  \country{Germany}
}
\affiliation{%
  \institution{TU Wien}
  \city{Vienna}
  \country{Austria}
}
\email{evropi.stefanidi@tuwien.ac.at}

\author{Pawe{\l} W. Wo\'zniak}
\orcid{0000-0003-3670-1813}
\affiliation{%
  \institution{TU Wien}
  \city{Vienna}
  \country{Austria}
}
\email{pawel.wozniak@tuwien.ac.at}

\author{Rainer Malaka}
\orcid{0000-0001-6463-4828}
\affiliation{
    \institution{University of Bremen}
    \city{Bremen}
    \country{Germany}
}
\email{malaka@tzi.de}

\author{Jasmin Niess}
\orcid{0000-0003-3529-0653}
\affiliation{%
  \institution{University of Oslo}
  \city{Oslo}
  \country{Norway}
  \postcode{}
}
\email{jasminni@ifi.uio.no}

\renewcommand{\shortauthors}{Mildner, et al.}

\begin{acronym}
\acro{HCI}{Human-Computer Interaction}
\end{acronym}

\begin{abstract}

Dark patterns are deceptive strategies that recent work in human-computer interaction (HCI) has captured throughout digital domains, including social networking sites (SNSs). While research has identified difficulties among people to recognise dark patterns effectively, few studies consider vulnerable populations and their experience in this regard, including people with attention deficit hyperactivity disorder (ADHD), who may be especially susceptible to attention-grabbing tricks. Based on an interactive web study with 135 participants, we investigate SNS users' ability to recognise and avoid dark patterns by comparing results from participants with and without ADHD. In line with prior work, we noticed overall low recognition of dark patterns with no significant differences between the two groups. Yet, ADHD individuals were able to avoid specific dark patterns more often. Our results advance previous work by understanding dark patterns in a realistic environment and offer insights into their effect on vulnerable populations.
\end{abstract}

\ccsdesc[500]{Human-centered computing~Empirical studies in HCI}
\ccsdesc[100]{Human-centered computing~HCI theory, concepts and models}
\ccsdesc[300]{Human-centered computing~Empirical studies in interaction design}
\ccsdesc[100]{Human-centered computing~Interaction design theory, concepts and paradigms}

\keywords{dark patterns, deceptive design, SNS, social media, ADHD, vulnerable populations, user study, quantitative study}

\maketitle
\section{Introduction}
Accelerated by data-driven incentives and powered by surveillance capitalism~\citet{zuboff_surveillance_2023}, unethical design has manifested in various digital technologies and user interfaces and drawn attention from researchers of the human-computer interaction (HCI) community for more than a decade~\cite{brignull_2022_wayback_types}. Described as deceptive design or ``dark patterns''~\cite{brignull_2022_wayback_types,gray_ontology_2023}\footnote{We are aware of a recent choice by the ACM Diversity, Equity, and Inclusion Council to categorise the term ``dark pattern'' as problematic~\cite{acm_words_2023}. However, we opted to use the term to stay in line with related work describing unethical and harmful UI practices. We acknowledge that the term could be misinterpreted with a connotation that ``dark'' could refer to ``bad'' or ``evil' intentions. However, we argue that the term makes a reference to \textit{hidden} or \textit{obfuscated} consequences for users~\cite{brignull_2022_wayback_types}, where intentions play a secondary role.}, over a decade of research in HCI has landscaped unethical design in various environments, including, but not limited to, e-commerce~\cite{mathur_dark_2019}, games~\cite{zagal_dark_2013}, and social networking sites (SNS)~\cite{mildner_ethical_2021}. As a medium connecting billions across the globe~\cite{sinclair2017facebook}, SNSs offer their users easy-to-use platforms to engage with others and (their) content but have been repeatedly the recipient of critique -- both from research~\cite{sina_social_2022,schaffner_understanding_2022} and regulatory actions~\cite{european_commission_commission_nodate} -- for relying on harmful design strategies, including dark patterns, and negatively affecting their users' well-being~\cite{ahn2013social,lin_why_2011}. 
In the context of SNSs, particularly problematic patterns are attention-grabbing or engaging dark patterns, aiming to maintain and increase user engagement through related design strategies, including auto-play or gamification mechanisms that many users may find difficult to avoid~\cite{monge_roffarello_defining_2023,mildner_defending_2023}. 

While related work has contributed efforts to investigate problems and call out issues in these regards, further research is needed, particularly with a focus on specific populations. This is especially the case for investigating and identifying any unique needs and interactions in the context of SNSs, including deployed dark patterns, given their wide adoption~\cite{mildner_about_2023,mathur_dark_2019,di_geronimo_ui_2020}. As both attention-grabbing and engaging elements are common dark patterns used in SNSs~\cite{monge_roffarello_defining_2023,mildner_about_2023,mildner_finding_2024} and traits of attention deficit hyperactivity disorder (ADHD)\footnote{In order to show respect for the different views and preferences communities and ADHD individuals have expressed regarding the use of person-first language, and similar to prior work (e.g.~\cite{spiel2022adhd, silva2023unpacking, stefanidi2024moodgems}), we use both ``ADHD individuals'' and ``individuals with ADHD''.} are associated with increasing amount of time on smartphones, 
including SNSs~\cite{aydin_trait-level_2024}, it becomes important to investigate this specific population, and whether this diagnosis can lead to particular challenges with manoeuvring these dark patterns. This need is underscored by a high prevalence of ADHD~\cite{polanczyk2007worldwide} and the fact that existing research indicates that individuals with ADHD are particularly susceptible to problematic technology use. For example, they are more likely to exhibit addictive behaviours in relation to video games~\cite{masi2021video}, a context where various dark patterns have been noticed~\cite{zagal_dark_2013}. Both HCI and dark pattern communities have repeatedly called for more attention to specific and vulnerable populations affected by unethical and harmful designs. 

To address this research gap and understand how vulnerable populations, specifically individuals with ADHD, experience dark patterns on SNSs, we designed a comparative, 2$\times$2 multi-factorial user study adopting similar measures as \citet{mildner_defending_2023}, who conducted an online survey to study SNS users ability to recognise dark patterns based on static images. However, research conducted by \citet{mildner_about_2023}, \citet{di_geronimo_ui_2020}, and \citet{gray_dark_2021} independently suggested a need to explore temporal aspects of dark patterns to better understand how they interact as users' rarely face instances in isolated contexts. To gain a more detailed view of our participants' ability to both recognise and avoid dark patterns in a more realistic setup, we developed a mock SNS in the form of a web application containing different types of dark patterns, allowing us to study dark patterns within user experiences and interactions instead of static screenshots. To consider suitable dark patterns for this study, we drew from \citet{mildner_about_2023} and \citet{monge_roffarello_defining_2023}, who described instances in the context of popular SNSs, as well as the context agnostic dark pattern ontology by \citet{gray_ontology_2023}, comprising high-, meso-, and low-level strategies. Through this approach, we seek to answer the following research questions: 

\begin{itemize}
    \item[\textbf{RQ1:}] Are there differences in how people with ADHD \textit{recognise} dark patterns on SNSs compared to people without ADHD?
    \item[\textbf{RQ2:}] Are there differences in how people with ADHD \textit{avoid} dark patterns on SNSs compared to people without ADHD?
\end{itemize}

Through our study design, tailored around understanding dark patterns in SNSs, this work advances recent work through three novel contributions: (1) We gain specific insights into how people with ADHD experience dark patterns on SNSs compared to people without ADHD, showing similar recognition capabilities overall, but also indications for different avoidance of dark patterns, depending on the context; (2) we expand previous work that considered dark patterns as isolated screen captures or other still images as their source by conducting an interactive study through a web application containing dark patterns that mirror realistic occurrences of dark patterns; 
and (3), we host and maintain our customisable Vue-based web application on \href{https://github.com/ThomasMildner/SNS-research/tree/CHI-2025}{\textcolor{blue}{\texttt{GitHub}}}\footnote{The source code used in this project's study is released as version v1.0.0 at \textcolor{blue}{\url{https://github.com/ThomasMildner/SNS-research/tree/CHI-2025}}, and archived under \textcolor{blue}{\url{https://doi.org/10.5281/zenodo.14761223}}.}~\cite{mildner_thomasmildnersns-research_2025}, including detailed descriptions and instructions, to support future, SNS-based user studies as these platforms become increasingly difficult to study in this regard.

\section{Background \& Related Work}
By studying the ability to recognise and avoid dark patterns in SNSs in consideration of vulnerable populations with ADHD, our research lies at the intersection between enhancing work on dark patterns and addressing the lack of understanding in HCI how they impact people with ADHD. Reflecting on this intersection, here we present a background of relevant efforts, while detailing related work more closely. This section begins with an overview of dark pattern scholarship before it dives into ADHD-related studies in HCI.

\subsection{Background on Dark Patterns}
After Harry Brignull coined the term ``dark pattern'' in 2010~\cite{brignull_2022_wayback_types}, research on the matter has accumulated a growing understanding of these practices and how they manifest in different digital technologies. 
Alongside this joined and trans-disciplinary effort, often combining HCI and legal perspectives (e.g.~\cite{gray_dark_2021}), a discourse emerged around the correct term to be used when referring to unethical design that meets the criteria of Brignull's original term, but without any potential racial implications~\cite{acm_words_2023}. While alternatives, such as deceptive design~\cite{sinders_whats_2022}, damaging design patterns~\cite{monge_roffarello_defining_2023} or manipulative design~\cite{sanchez_chamorro_my_2024}, have been used by different authors in this context, no agreement has been established, while voices within the community have reiterated over the connection of dark to hidden, not evil, practices~\cite{obi_lets_2022}, contradicting claims of racial references. For the sake of consistency with other work from regulation and academic research, we will remain using \textit{dark patterns}, while referring to hidden consequences and not evil practices. 

Meanwhile, dark patterns have been catalogued more agnostic to specific domains~\cite{gray_dark_2018} or in more context-specific environments like e-commerce~\cite{mathur_dark_2019,gray_dark_2018}, mobile applications~\cite{di_geronimo_ui_2020, gray_dark_2018}, and conversational user interfaces~\cite{owens_exploring_2022, mildner_listening_2024}.
A recent contribution in this regard is the synthesis of various such streams into a shared ontology~\cite{gray_ontology_2023}, aiming to facilitate continuing work by providing a shared vocabulary. The ontology assessed over 200 dark patterns and identified 64 individual types across a three-tier hierarchy, including context and technology-agnostic high- and meso-level strategies, as well as low-level types that are context-specific. The work draws from sources including regulatory reports as well as academic research. Reports stem, for instance, from the OECD~\cite{oecd__2022_dark} or the FTC~\cite{FTC2022}, describing how online users are frequently exposed to harmful design tricks that require further attention to ensure user safety. In academia, on the other hand, the ontology builds on key contributions such as \citet{mathur_dark_2019}, who identified and described dark patterns in a large data set of e-commerce websites. In this work, \citet{mathur_dark_2019} further proposes a framework organising dark patterns between five characteristics, constituting an additional contribution to the field. Later, \citet{mathur_what_2021} introduced a sixth characteristic to include the gaming dark patterns by \citet{zagal_dark_2013}. Therefore, the final six characteristics are: (1) asymmetric, (2) covert, (3) deceptive, (4) hides information, (5) restrictive, and (6) disparate treatment. Our study design adopts these characteristics to assess our participants' abilities to recognise and avoid dark patterns, following the approach by \citet{mildner_defending_2023}.

\subsection{Dark Patterns in Social Media}
Unlike other commercial domains, dark patterns in SNSs manifest not primarily as tricks to manipulate or coerce users into overspending money but time~\cite{mildner_ethical_2021} or personal data~\cite{bosch_2016_privacy}. With a focus on this context, \citet{bosch_2016_privacy} described a variety of privacy-infringing dark patterns, including \textit{Privacy Zuckering} and \textit{Bad Defaults}, which in themselves describe tricks to steer users into sharing sensible data against their interests, but pose an even bigger threat to users' privacy when combined~\citet{mildner_defending_2023}. \citet{mildner_about_2023} analysed recorded walkthroughs of four SNSs, Facebook, Instagram, TikTok, and Twitter (before the platform was re-branded to X). Based on a taxonomy of 80 dark pattern types, the study noticed 44 instances throughout the platforms. Furthermore, the authors noticed SNS-specific types within the engaging and governing strategies. The first restricts users' agency to maintain their time spent on SNSs in line with their goals, while the latter hinders them from reaching their goals by obstructing navigational aspects of the user interface.

Similar to engaging strategies, \citet{monge_roffarello_defining_2023} defined attention-grabbing dark patterns, thereby building on prior work by \citet{lukoff_how_2021}. In their work, the authors conducted a systematic literature review to identify instances in related work that meet the criteria of design tricks capturing their audience's attention, potentially steering them away from their original goals, and leading to a loss-of-control feeling. In a similar vein, \citet{schaffner_understanding_2022} studied SNS users' ability to delete their accounts, noticing an important deficit in this regard. Design mechanisms hindered users from achieving their goals or kept them in a false state of belief that they deleted their accounts when they actually did not.

Dark patterns specific to SNSs seem to follow different incentives compared to others deployed in other domains, as they have to keep users' satisfaction high while guiding users' choices towards the service provider's goals~\cite{mildner_ethical_2021}. The presence of designs that keep users active on platforms by manipulating their choice architecture, for instance, through auto-play mechanisms, is concerning. Especially, as work has repeatedly noticed a lack in people's ability to recognise dark patterns sufficiently and effectively avoid them~\cite{maier_dark_2020,bongard-blanchy_i_2021,di_geronimo_ui_2020}. In a recent study, \citet{mildner_defending_2023} replicated parts of a related study by \citet{di_geronimo_ui_2020} to show similar effects in the context of SNSs. While these studies mainly deployed static images of dark patterns or, in the case of \citet{di_geronimo_ui_2020}, recorded videos, our study design expands on these works by implementing an interactive web-based study for SNSs. Thus, we gain more detailed insights into our participants' engagement with interfaces containing dark patterns. 

\subsection{Attention Deficit Hyperactivity Disorder (ADHD)}
With the implications of dark patterns being serious for any person, they may be even more severe for individuals with attention deficit hyperactivity disorder (ADHD)~\cite{adhd_ICD}, who may be differently affected by engaging and attention-grabbing strategies.
ADHD is the prevailing neurodevelopmental diagnosis in children~\cite{peroumental2013, zasler2013neurobehavioral}, and continues into adulthood~\cite{wender2001adults}.
Deficits in "executive function" are often proposed as a potential cause of ADHD or at least as a significant factor in its disruption~\cite{furman2005attention}. Although definitions of executive functioning vary among researchers, it generally encompasses skills such as self-monitoring, maintaining focus despite distractions, and effective self-organization, among others~\cite{furman2005attention}.
ADHD individuals can exhibit diverse behaviours based on the presentation type: predominantly inattentive, hyperactive-impulsive, or combined presentation~\cite{gregg2000definition, american2013diagnostic}, depending on which of these characteristics is prevalent. 
ADHD has been associated with academic underachievement, bedtime resistance, disruptive behaviours, poor self-regulation of emotions, and social difficulties, such as issues in interacting with peers~\cite{wehmeier2010social, faraone2015attention, sonne2015designing}.
Despite these challenges that have been associated with ADHD, literature has been adopting and advocating for a more strengths-based approach to ADHD (e.g.~\cite{schippers2024associations, charabin2023m, mahdi2017international, sedgwick2019positive}). Researchers, in particular, have found a positive correlation between ADHD and creativity~\cite{mahdi2017international, white2006uninhibited, hoogman2020creativity}, as well as with cognitive flexibility~\cite{schippers2024associations} and hyperfocus~\cite{schippers2024associations, hupfeld2019living}.

The HCI community has a long tradition of supporting vulnerable populations, such as those with ADHD. In the last decade, HCI has increasingly recognised the responsibility that digital technologies have great potential to both help and harm marginalised groups~\cite{dalton2013neurodiversity}. The interest in researching neurodivergent groups, more specifically individuals with ADHD, is highlighted by both the number of literature reviews that have been published on the topic (e.g.~\cite{sonne2016assistive, cibrian2020research, cibrian2022potential, stefanidi2022designing}), as well as by the number of interactive technologies (e.g.~\cite{cho2002attention, pina2014situ, sonne2016changing, stefanidi2024moodgems}) that have been designed aiming to support them.


While recent efforts have investigated dark patterns with certain vulnerable populations, such as children~\cite{renaud2024we} or teenagers~\cite{sanchez_chamorro_my_2024}, there is remains a need to explore how dark patterns affect a specific vulnerable population, i.e. individuals with ADHD. This is supported by both the prevalence of ADHD~\cite{polanczyk2007worldwide} and by research that has demonstrated their vulnerability to problematic technology use~\cite{werling2022problematic}, for instance exhibiting more addictive behaviours with respect to video games than their non-ADHD peers~\cite{masi2021video}.
This work addresses the above by investigating users' experience with dark patterns in SNSs by comparing the interactions of ADHD individuals to those of non-ADHD individuals.



\section{Methodology}
To understand how people with ADHD are affected by dark patterns in SNSs, we designed a comparative, web-based study with a mock SNS. We designed a 2$\times$2-factorial study, allowing us to address our research questions about the differences in recognition and avoidance between ADHD individuals and people without ADHD. Here, we describe the individual components of our methodology.

\begin{figure*}[t!]
    \centering
    \includegraphics[width=.7\textwidth]{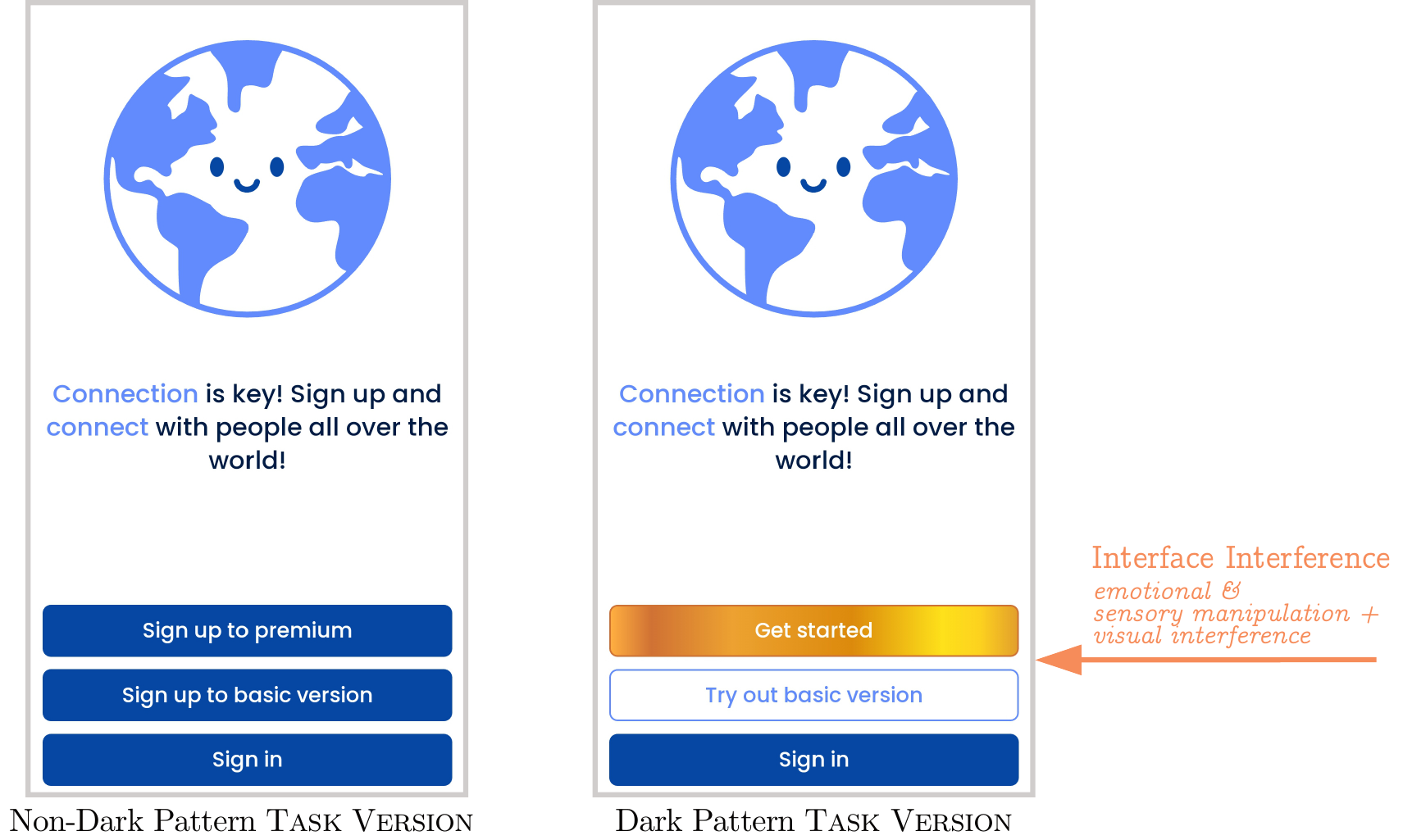}
    \caption{This figure shows the ``Choose Plan'' segment of Task 1. On the left, the non-dark pattern version does not emphasise specific interface elements. On the right is the dark-pattern version, which utilises \textit{Interface Interference} dark patterns to visually highlight the premium account option over the basic alternative.}
    \Description[Example screenshots from Task 1.]{This figure demonstrates the "Choose Plan" segment of Task 1. On the left, the non-dark pattern version is shown, not emphasising specific interface elements. On the right is the dark-pattern version, which utilises Interface Interference dark patterns to visually highlight the premium account option over the basic alternative.}
    \label{fig:example-t1}
\end{figure*}

\begin{figure*}[t!]
    \centering
    \includegraphics[width=.7\textwidth]{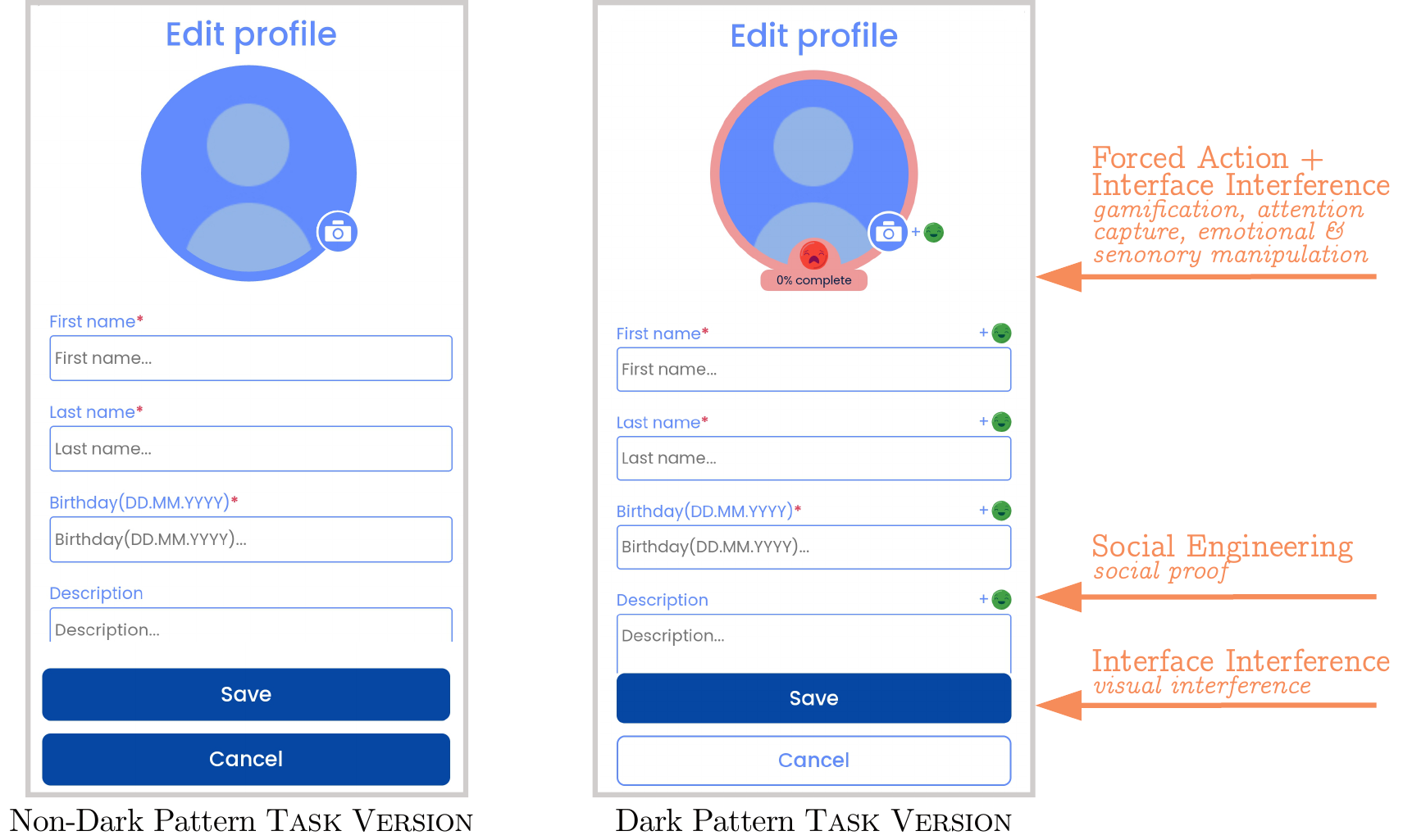}
    \caption{This figure shows the ``Edit Profile'' segment of Task 2. On the left, the non-dark pattern version does not emphasise options for users to enter data. On the right, the dark-pattern version utilises multiple instances of \textit{Interface Interference} dark patterns to visually highlight the save button and uses \textit{Social Engineering} and \textit{Forced Action} patterns to steer the user into sharing more personal data, for instance through a smiley that turns happier the more data is entered.}
    \Description[Example screenshots from Task 2.]{This figure demonstrates the "Edit Profile" segment of Task 2. On the left, the non-dark pattern version shows no emphasised options for users to enter data. On the right, the dark-pattern version utilises multiple instances of Interface Interference dark patterns to visually highlight the save button and uses Social Engineering and Forced Action patterns to steer the user into sharing more personal data, for instance, through a smiley that turns happy the more data is entered.}
    \label{fig:example-t2}
\end{figure*}

\subsection{Web Application}

Users are likely to face different types of dark patterns both within the same interaction as well as throughout their journeys to reach their goals. In this regard, \citet{gray_temporal_2023} raised the importance of considering temporal aspects of dark patterns when analysing them in interfaces. 
Reflecting on their realistic appearance in interactive environments, we opted for developing a functional mock SNS platform in the form of a web application. An important benefit of this choice is the ability to record our participants' choices and track their journey throughout given tasks -- something that is increasingly difficult on current SNSs as API prices surge and UI elements are implemented to obfuscate such functionalities, hindering research in these regards. 

To this end, we developed a web application, based on \texttt{Vue.js}, which is hosted on \href{https://github.com/ThomasMildner/SNS-research/tree/CHI-2025}{\textcolor{blue}{\texttt{GitHub}}}~\cite{mildner_thomasmildnersns-research_2025}. The application automatically records participants' interactions with individual UI elements while tracking their choices and paths they have taken. To study dark patterns on SNSs,  \citet{mildner_defending_2023} chose an online survey study design with SNS-based screenshots that did or did not contain dark patterns. While their study proved effective in identifying differences, it is limited to only observing static dark patterns. Our approach aims to mitigate this limitation and allows us to collect all needed data to understand the temporality of dark patterns further, allowing us to compare how different user groups navigate SNSs and experience dark patterns. We designed the web application to be relatively easy to customise for individual purposes of future studies. Figures~\ref{fig:example-t1} and \ref{fig:example-t2} showcase example screenshots that demonstrate differences between versions of the same interface, highlighting included dark patterns.

\begin{figure*}[t!]
    \centering
    \includegraphics[width=\textwidth]{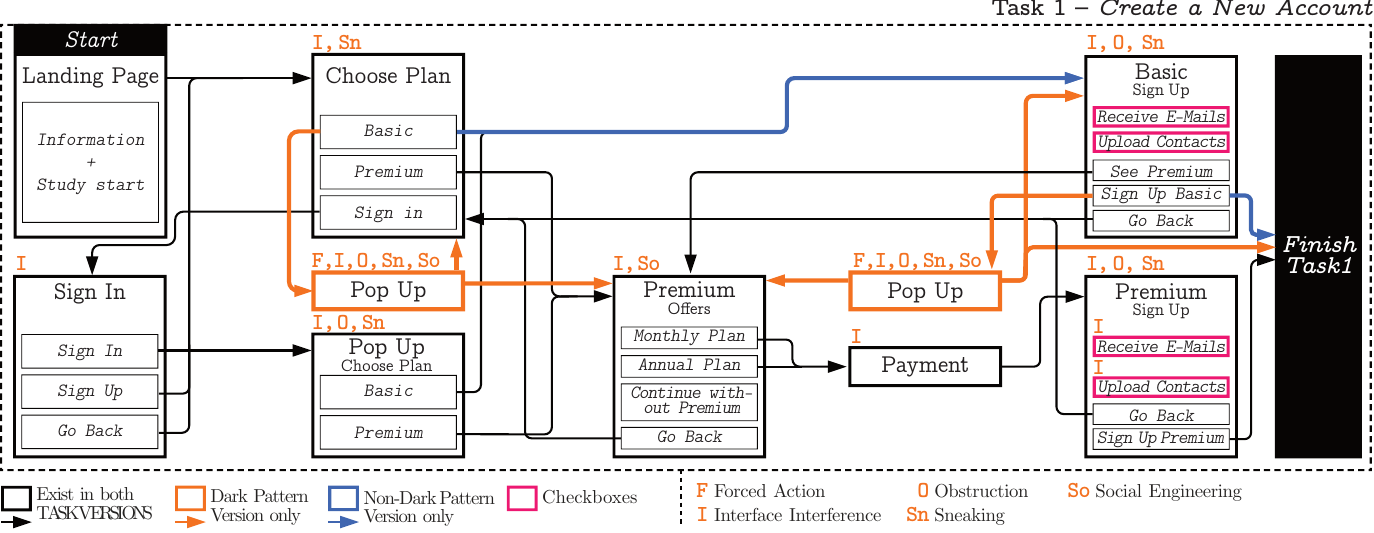}
    \caption{This figure illustrates the task flow of Task 1. Both \textsc{Task Versions} are integrated, although distinguished through different outlines and arrow colours. Orange indicates present paths and options only present in the dark pattern \textsc{Task Version}, while blue only exists in the neutral \textsc{Task Version}. Magenta outlines three included checkbox options. Furthermore, present dark patterns are implied through letters representing high-level strategies from \citet{gray_building_2024}. Colours and symbols are explained in the legend below the diagram.}
    \Description[Task flow diagram for Task 1.]{This figure illustrates the task flow of Task 1 using a wireframe layout. Both Task Versions are integrated, although distinguished through different outlines and arrow colours. Orange indicates present paths and options only present in the dark pattern Task Version, while blue only exists in the neutral Task Version. Magenta outlines three included checkbox options. Furthermore, present dark patterns are implied through letters representing high-level strategies from Gray et al.'s 2024 ontology.}
    \label{fig:task1-flow-diagram}
\end{figure*}

\subsection{Apparatus}
\begin{figure*}[t!]
    \centering
    \includegraphics[width=\textwidth]{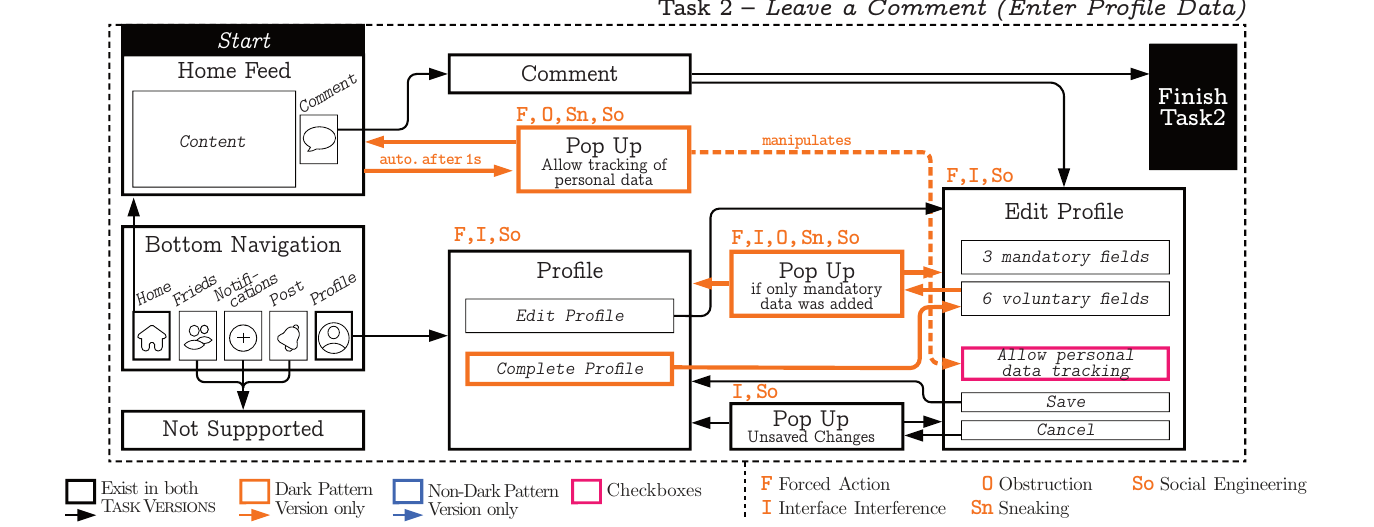}
    \caption{This figure follows the task flow of Task 2. As in Figure~\ref{fig:task1-flow-diagram}, non-dark pattern and dark pattern \textsc{Task Versions} are integrated, while different outlines and arrow colours indicate changes between them. Orange indicates present paths and options only present in the dark pattern \textsc{Task Version}, while blue only exists in the neutral \textsc{Task Version}. Magenta outlines three included checkbox options. Furthermore, present dark patterns are implied through letters representing high-level strategies from \citet{gray_building_2024}. Colours and symbols are explained in the legend below the diagram.}
    \Description[Task flow diagram for Task 1.]{This figure illustrates the task flow of Task 2 using a wireframe layout. Both Task Versions are integrated, although distinguished through different outlines and arrow colours. Orange indicates present paths and options only present in the dark pattern Task Version, while blue only exists in the neutral Task Version. Magenta outlines three included checkbox options. Furthermore, present dark patterns are implied through letters representing high-level strategies from Gray et al.'s 2024 ontology.}
    \label{fig:task2-flow-diagram}
\end{figure*}

Based on our web application, we implemented two sequential SNS-related tasks for our participants to complete on their mobile phones. The first required them to create a new account; the second prompted them to comment on an existing post. We acknowledge that this choice can introduce an order bias, potentially resulting in a learning effect. However, we opted for this particular study design to allow the utilisation of a scenario benefitting comparability between participant data while accommodating common SNS experiences. This tradeoff is further enhanced by the number of selected participants to understand any differences between our independent variables. In our 2$\times$2 factorial design, we designed each task to be either neutral in the display of information and choices or utilise dark patterns to coerce users. For the neutral task versions, we drew from work on responsible~\cite{leimstadtner_investigating_2023, hansen_nudge_2013} and value-centred design~\cite{friedman_value_2013} concepts. For the dark pattern exposing tasks, we adopted engaging and governing strategies by \citet{mildner_about_2023} as well as attention-grabbing types from \citet{monge_roffarello_defining_2023}, while utilising the dark pattern ontology~\citet{gray_ontology_2023} to choose specific types. Below, we describe the two tasks of our web application in closer detail. Because the web application is too complex to be fully described here, we include task flows in Figure~\ref{fig:task1-flow-diagram} and \ref{fig:task2-flow-diagram}, while adding screenshots of all sections and task versions to this paper's supplementary material for closer inspection.

\paragraph{Task 1}
When entering the web application, in the first step, participants will be informed about the task for which they have to create a new account. While we aimed to keep the non-dark pattern version of our mock SNS as neutral as possible in terms of persuasive and deceptive design, we integrated various dark patterns in the respective version. To this end, we drew from real SNSs, including Facebook and TikTok, but also from Gray et al.'s ontology~\cite{gray_building_2024} as well as the studies by \citet{monge_roffarello_defining_2023} and \citet{mildner_about_2023} to align our application closely with their research on attention-grabbing and engaging effects. 
To solve the task, participants have the option to choose between basic and premium accounts, which require similar steps to sign up. 
In the dark pattern version, dark patterns like \textit{Interface Interfances}, \textit{Sneaking}, or \textit{Social Engineering} try to enforce a premium membership in various instances.
Once they successfully create their account, participants automatically continue with Task 2, which begins in the app's home screen.
Figure~\ref{fig:task1-flow-diagram} follows the entire user journey throughout this task, highlighting differences between neutral and dark pattern versions.

\paragraph{Task 2}
Task 2 requires participants to comment on an existing post. As with Task 1, multiple dark patterns aim to steer and manipulate our participants' choices in the dark pattern version. This includes an instance similar to TikTok, where app users face a running video before a prompt asks them to agree to their personal data being tracked. \citet{mildner_about_2023} described this particular dark pattern as \textit{Decision uncertainty}, as part of their \textit{Engaging Strategies} when users' focus is overloaded through additional, attention-grabbing mechanisms. 
Before they are able to comment, however, the web application prompts them to first add certain data to their profile. 
In total, they can add nine things: (1) a profile image (a mock-image loader with a small selection of pre-loaded images was included in the web application), (2), first name, (3) family name, (4), date of birth, (5), a profile description, (6) gender, (7) phone number, (8) city, and (9) job title. Importantly, only items 2-4 are mandatory to comment on the post and finish Task 2 in the non-dark pattern version, while participants had to add data for items 2-4 as well as 6 and 7 in the dark pattern version. The other fields were optional. As visualised in Figure~\ref{fig:example-t2}, we designed the dark pattern version to coerce participants into sharing more data than possibly desired. 

\subsection{Pre-study}
To ensure that our study would produce meaningful insights, we conducted a pre-study with 14 participants (female = 4, male = 6, non-binary = 2, preferred not to say = 1, other = 1) through convenience sampling. Their age ranged between 18 to 31 years, with an average of $M=23.7$ ($SD=3.77$). After providing the necessary information and gaining their consent, seven participants reported having ADHD, with the other seven stating they did not. We asked participants of this pre-study to complete both tasks to identify any possible errors and to collect their feedback based on the presentation of the tasks. This feedback led to minor changes and improvements in the study setup, mainly to mitigate compatibility issues between different devices.

\subsection{Procedure}
For the main study, we recruited participants from the online recruiting platform Prolific~\cite{prolific} to ensure fair compensation for our participants' time and efforts. Following our study design, we recruited two groups of participants for active SNS users with and without ADHD, using Prolifics' available pre-screening criteria. Participants of either group were excluded from seeing the other version respectively. Through Prolific, we required participants to have a desktop and a mobile device to complete the study. In a first step, they were forwarded to a survey on ScoSci Survey~\cite{leiner2024soscisurvey}, which they should access via their desktop device, informing participants about the study's nature and requesting their consent.

\begin{figure*}[t!]
    \centering
    \includegraphics[width=\textwidth]{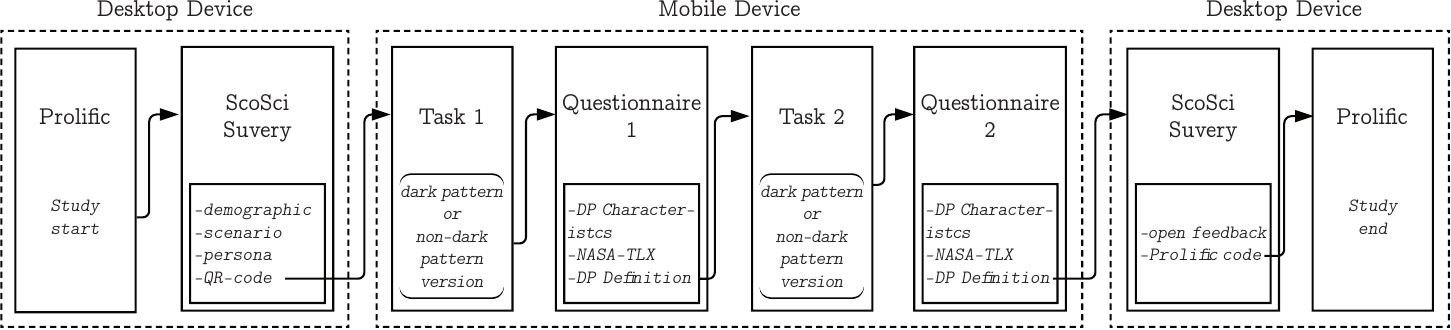}
    \caption{This flowchart shows the procedure of our experimental setup. Starting on Prolific from a desktop device, participants are forwarded to ScoSci Survey, where they are asked to enter demographic data and see scenario and persona information before a QR code links them to the two tasks on the web application on their mobile device. After each task, we asked participants to complete the same questionnaires before they were prompted to return to ScoSci Survey to receive a completion code that allowed them to finish the study on Prolific.}
    \Description[Flowchart for the study procedure.]{This figure shows the procedure of our experimental setup. Starting on Prolific from a desktop device, participants are forwarded to ScoSci Survey, where they are asked to enter demographic data and see scenario and persona information before a QR code links them to the two tasks on the web application on their mobile device. After each task, we asked participants to complete the same questionnaires before they were prompted to return to ScoSci Survey to receive a completion code that allowed them to finish the study on Prolific.}
    \label{fig:study-flow}
\end{figure*}

Participants were presented with a scenario, including a persona, to adopt when completing the tasks. We equipped the persona with plenty of personal data that participants could enter when asked, protecting their privacy. Moreover, relying on a scenario and persona helped us mitigate personal choices influenced by individual circumstances and increase the comparability of results. As the scenario and persona remained on the desktop device, participants could quickly access the information throughout the study. Both scenario and persona are included in Appendix~\ref{app:survey-material}. After they read the information, participants were provided with a QR code which they could activate via the mobile device to load a specific permutation of the study's versions. 

On entering the web application, participants were once again informed about the study with a focus on individual tasks. After completing each task, the web application included a set of questionnaires participants were to answer. Finally, after the second task, the web application would redirect participants to the survey, where they were provided with a code to redeem their compensation on Prolific. Figure~\ref{fig:study-flow} offers an overview of the study procedure.

\subsection{Conditions}
Our study manipulates two factors: \textsc{Task Version} and \textsc{Group Type}.
The \textsc{Task Version} factor refers to whether the interface used in the task included \textit{dark patterns} (manipulative design techniques intended to deceive or mislead users) or was a \textit{non-dark pattern} version, which lacked such manipulative elements.

The \textsc{Group Type} factor refers to the participants' self-reported status regarding ADHD, with levels consisting of participants who identified as having \textit{ADHD} and those who identified as \textit{non-ADHD}. These terms will be used consistently when reporting effects across these factors for clarity.

\subsection{Measures}

\input{tables/mathur_characteristics}

For this study, we drew inspiration from \citet{mildner_defending_2023} who studied SNS users' ability to recognise dark patterns based on screenshots sampled from Facebook, Instagram, TikTok, and Twitter (before it was rebranded to ``X''). In this study, the authors used questions based on Mathur et al.'s dark pattern characteristics~\cite{mathur_dark_2019,mathur_what_2021} as well as given dark pattern definition~\cite{mathur_dark_2019}. As seen in Table~\ref{tab:characteristics_suvey}, we changed the questions into statements for better comprehension. Further, we included the later added sixth characteristic, \textit{Disparate Treatment}, which was not yet included in Mildner et al.'s\cite{mildner_defending_2023} work. Per task, we asked our participants to rate the statements using a 7-point Likert scale (``Not at all'' - ``Extremely'').

To better understand how dark patterns in SNSs may affect cognitive load between individuals with and without ADHD, we included the NASA-TLX questionnaire~\cite{hart1988development}, also using a 7-point Likert scale. At the end of each question set, similar to \citet{mildner_defending_2023}, we used Mathur et al.'s~\cite{mathur_dark_2019} definition to ask our participants, per task, whether they noticed any design elements meeting this description, using ``yes'', ``no'', ``maybe'' response types.


\input{tables/task-permutation}

\subsection{Participants}
To determine the number of participants required to gain meaningful insights, we conducted an \textit{a priori} power analysis with G*Power~\cite{faul2009gpower}. As input parameters, we assumed a mid-sized effect size of $d=0.5$ and a power of $\beta=0.8$. Hence, the analysis suggests 64 participants per group or a total of 128 participants. Here, we describe the recruitment of both participant groups and their demographics.

To conduct the study, we relied on the online platform Prolific~\cite{prolific} for recruiting participants. We increased the number of each group by $\sim$10\% (13 participants) following past experiences of participants retracting their data, unfinished responses, or otherwise defective or incomplete data. Finally, we recruited 144 participants to conduct the study with 72 ADHD and 72 non-ADHD individuals. 
Before the study, we reached out to Prolific and learnt that, for ethical reasons, participants in the ADHD group did not have to submit official diagnoses to confirm having ADHD. Instead, participants self-reported having ADHD. However, they were asked whether a medical or psychological professional had given them a diagnosis of ADHD on a voluntary basis. All participants had to be active users of an SNS platform at least on a weekly basis and showed similar experiences with almost daily SNS usage. On the one hand, participants from the ADHD group used SNSs at an average of $M=6.72$ days per week ($SD=1.03$). On the other, the non-ADHD group used SNSs $M=6.79$ days ($SD=0.66$) a week on average.
We designed the study to be completed within 20 minutes for an hourly payment of at least £9 (£3 per participant for the given time). Ultimately, the median completion time was 13 minutes, resulting in an average hourly payment of £14,77 (£3,21 per participant). The data of nine participants had to be excluded from further analysis as the result of incomplete answers or failing Prolific's checks. Finally, we gained complete data from 135 participants, with 68 participants in the ADHD- and 67 in the non-ADHD group with 270 completed data sets for between groups and both tasks. Table~\ref{tab:task-permutation} gives an overview of our participants' distribution between task permutations and groups after removing incomplete data sets.



\begin{figure*}[t]
    \centering
    \includegraphics[width=\textwidth]{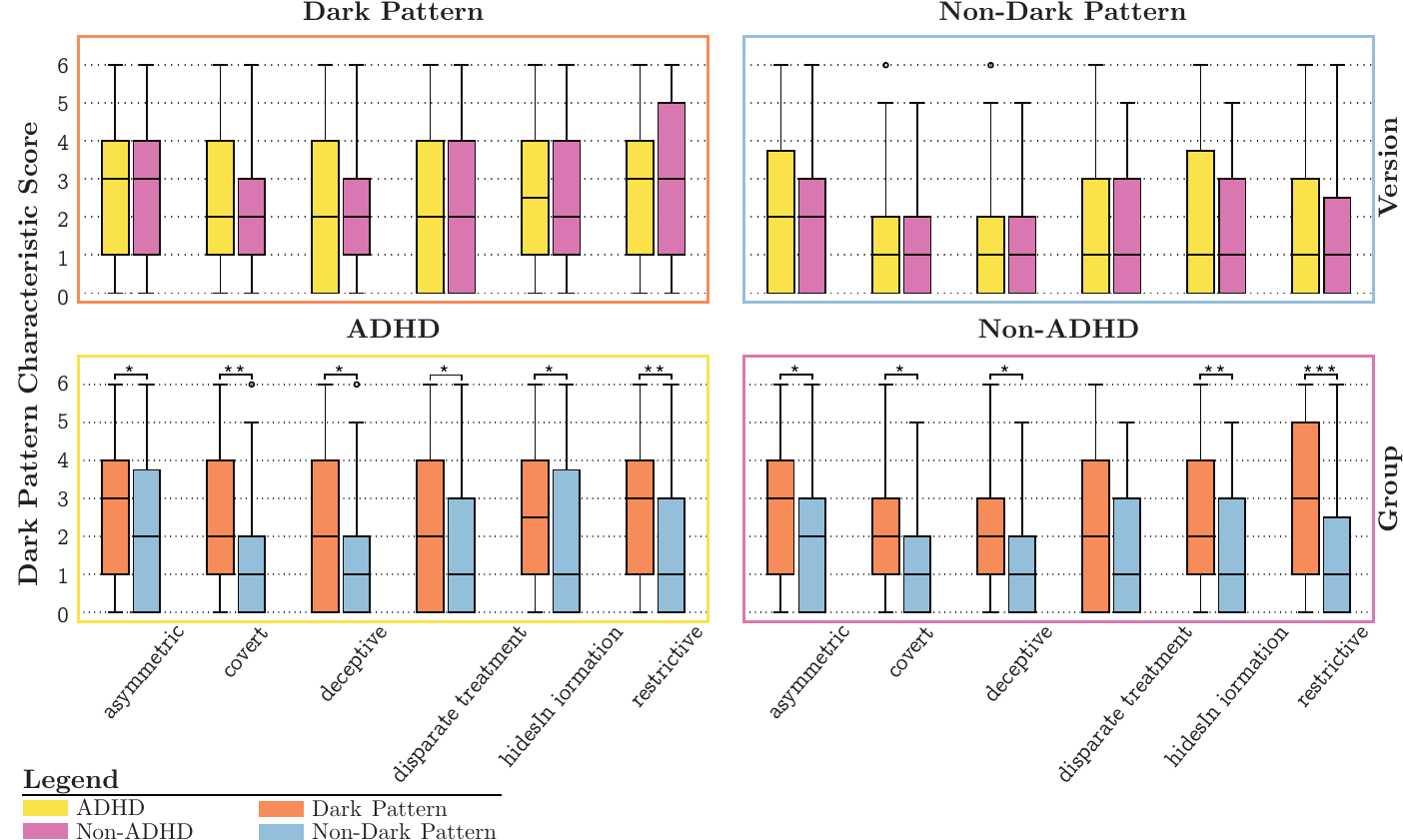}
    \caption{Accumulated results between \textsc{Task Versions} for the dark pattern characteristics of I1-I6. The four box plots show the results between the factors of \textsc{Task Versions} $\times$ \textsc{Group Type}. Colours are used to differentiate between factors mapped to the same colour scheme we used in the task flow diagrams in Figures~\ref{fig:task1-flow-diagram} and \ref{fig:task2-flow-diagram}.}
    \Description[Multiple bar plots of the dark pattern characteristics of items i1-16.]{This figure visualises the accumulated results between Task Versions for the dark pattern characteristics of I1-I6. The four bar plots show the results between the factors of Task Versions times Group Type. Colours are used to differentiate between factors. It can be seen that between the group types, no significant differences are noticeable, while dark patterns significantly affected either group type's decisions.}
    \label{fig:characteristics-results}
\end{figure*}

\section{Results for RQ1}
\input{tables/anova-results-q1}
The questionnaires offer interesting data for answering our first research question about dark pattern recognition. As the data consists of Likert scale responses and discrete counts, we adopt a non-parametric approach to analysing the data. To that end, we employ the aligned rank transform for ANOVA procedures. \citet{wobbock_align_2011} designed this method specifically for studies of the type conducted here. We checked cell densities and found that none exceeded 10, indicating that an ART can be safely conducted~\cite{luepsen_aligned_2017} and is more rigorous than a parametric approach~\cite{wobbock_align_2011}.

In this section, we report the results, beginning with our participants' demographics. We then look into the results for the dark pattern characteristics (items I1-I6 as shown in Table~\ref{tab:characteristics_suvey}), followed by our evaluation of participants' responses when given the definition (I7). Finally, we present our results of the NASA-TLX questionnaire to investigate the role of cognitive load when participants engage with SNS interfaces that deploy dark patterns. Because our overall results are too long to be presented in this paper, we will focus mainly on interesting findings while including all test results in this paper's supplementary material.

\subsection{Participant Demographics}
Of the participants in the ADHD group, 33 identified as female and 35 as male, with an average age of $M=$25.71 years ($SD = 5.29$). When asked about their highest education, 30 replied with holding a high-school diploma, 29 held a bachelor's degree, 5 a master's degree, and 4 replied with ``other''.
As regular SNS usage was a participation criterion, we asked participants about their usage. On average, participants with ADHD used SNSs on $M=$6.72 days per week ($SD=1.03$) and $M=$4.72 hours per day ($SD=3.11$).
At the time of conducting this experiment, participants of the ADHD group were predominantly located in South Africa (13), Poland (13), and Mexico (7), with additional representation from Portugal (7), Italy (5), Hungary (3), Greece (2), Latvia (3), Spain (2), United Kingdom (2), Chile (3), Belgium (1), Canada (1), Czech Republic (1), Denmark (1), Estonia (1), France (1), Germany (1), and the Netherlands (1).

In the non-ADHD group, 33 identified as female, 32 as male, 1 as non-binary, and 1 as gender-queer, with an average age of $M=$28.40 years ($SD=8.19$). Regarding highest education, 18 replied with high-school diplomas, 31 held a bachelor's degree, 12 a master's degree, 4 stated ``other'' forms of education while 2 did not disclose their highest education. Asked about their SNS weekly usage, non-ADHD participants used SNSs on $M=$6.79 days ($SD=0.66$) and $M=$4.18 hours per day ($SD=3.29$). Participants of this group were located in South Africa (22), followed by Italy (7) and Portugal (7), with further participants from the United Kingdom (3), Greece (3), Hungary (3), Germany (2), France (2), Chile (1), Croatia (1), Canada (1), Finland (1), Latvia (1), Mexico (1), Netherlands (1), New Zealand (1), Norway (1), Slovenia (1), South Korea (1), Spain (1), and Switzerland (1).


\subsection{Dark Pattern Characteristics -- Items I1-I6}

We conducted two-way repeated-measures ART-ANOVAs, separately for each of the two tasks, to test for the effect of \textsc{Task Version} and \textsc{Group Type} on the individual item scores for of the items I1--I6. Table~\ref{tab:anova_i1-7} and Figure~\ref{fig:characteristics-results} illustrate these results. We did not find any significant interaction effects. However, we found significant main effects of \textsc{Task Version} on all items for Task 1 and no main effects for \textsc{Group Type}. Interestingly, we found no significant main effects in Task 2.
These results imply that we cannot conclude whether or not the two participant groups differed in the ability to differentiate between the versions of the web applications in terms of dark pattern characteristics.

\subsection{Definition -- Item I7}
Using Mathur et al.'s~\cite{mathur_dark_2019} dark pattern definition, we asked participants after each task whether they noticed any matching interface elements using ``yes'', ``no'', or ``maybe'' response types. 
We conducted a one-way repeated-measures ART-ANOVA
to test the effect of \textsc{Task Version} and \textsc{Group Type} on whether participants noticed matching interface elements (I7). We found no significant effects. This is in line with results for items I1-I6.


\subsection{NASA-TLX Results}
\begin{figure*}[t!]
    \centering
    \includegraphics[width=\textwidth]{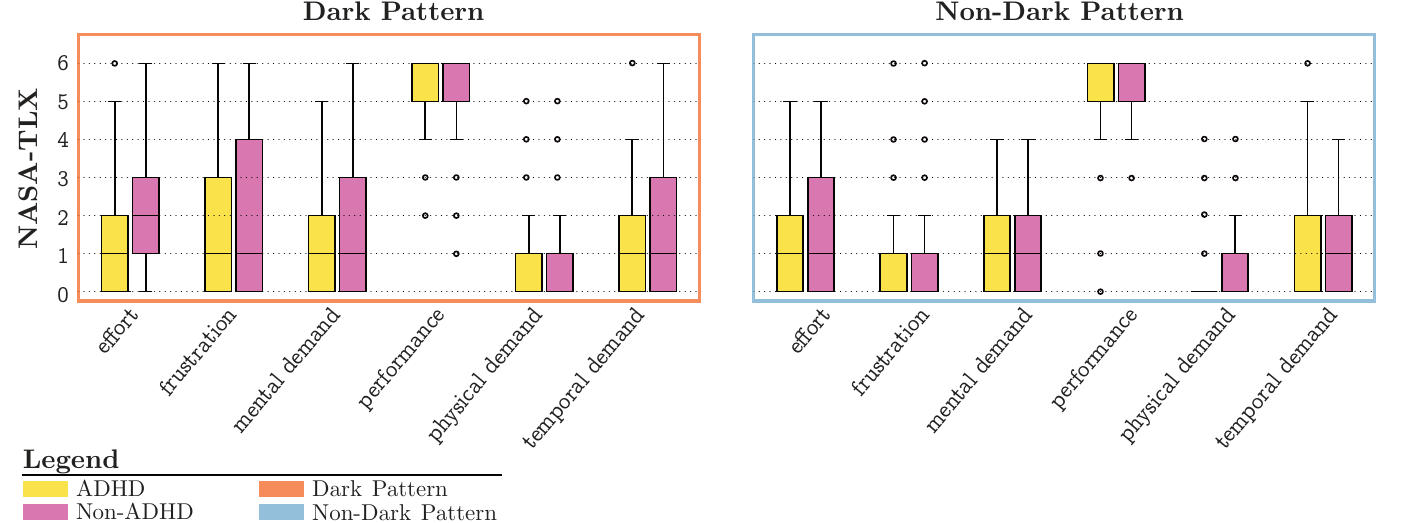}
    \caption{This figure visualises the accumulated results between \textsc{Task Versions} for the NASA-TLX. Colours are used to differentiate between factors.}
    \Description[Bar plots for the NASA-TLX]{This figure visualises the accumulated results between Task Versions for the NASA-TLX. Colours are used to differentiate between factors. For the dark pattern Task Version, effort, frustration, and mental demand were lower among participants from our ADHD group.}
    \label{fig:nasatlx-results}
\end{figure*}
\input{tables/anova-nasatlx}

We used the NASA-TLX to gain additional insights into how the presence of dark patterns may affect cognitive load. Moreover, it offers a validated and commonly used metric, unlike items I1-I7, contributing to the validity of the study at hand.
We conducted two-way repeated measures ANOVAs to test the effect of \textsc{Task Version} and \textsc{Group Type} on NASA-TLX subscales and the total scale score.  The presence of dark patterns had a significant effect on the total TLX score ($F(1, 114)=10.22; p=0.002$). However, the test did not show significant differences for \textsc{Group Type} nor an interaction effect between the \textsc{Task Version} $\times$ \textsc{Group Type}.

Looking at the subscales, we found significant main effects, shown in Table~\ref{tab:anova_nasa-tlx}. 
The subscales for effort, frustration, and physical demand show significant effects, suggesting that participants with ADHD experienced different cognitive load during our experiment. Looking at the individual scores given between the \textsc{Group Types}, we noticed that participants with ADHD generally reported lower scores for these subscales, as visualised in Figure~\ref{fig:nasatlx-results}.
Furthermore, \textsc{Task Version} containing dark patterns affected the mental demand and frustration of all participants.
These results show that participants with ADHD experienced lower effort, frustration, and physical demand during the experiment than our non-ADHD participants, while dark patterns increased the cognitive load for all.

\section{Results for RQ2}
To provide insights for our second research question about dark pattern avoidance, we analysed interaction data from our participants between \textsc{Task Versions}. In Task 1, we investigated which account type they chose between basic and premium. In Task 2, we were interested in the number of personal data entered. In addition to these two tasks, we included three checkboxes throughout both tasks, which would grant specific access or allow certain actions. Through dark patterns, such as \textit{Bad Default} and \textit{Interface Interference}, we manipulated the interface to potentially enforce compliance. 
Here, we report the results of additional ART-ANOVA tests for these choices and learn about our participants' ability to avoid dark patterns. Importantly, when we refer to our participants' choices, we do not imply agency in making an informed decision (which is hindered by the presence of dark patterns~\cite{mildner_dark_2024}) but rather the outcome of their interaction.

\subsection{Avoidance in Task 1---Premium Accounts}
In the first task, participants were tasked to create an account but were offered two options: One was a \textit{basic}, free-to-use, and the other represented a \textit{premium} account including monthly or yearly subscription fees, which we accumulated for the purpose of this study. 
Within the ADHD group, using the dark pattern version, 87\% selected the basic version. In the non-dark pattern version, participants of the ADHD group chose the basic version 91\% of the time. In the non-ADHD group, using the dark pattern version, only 67\% chose the basic version, whereas 97\% chose this option in the non-dark pattern task.
Conducting an ART-ANOVA on our participants' subscription choices shows significant interaction between \textsc{Task Version} $\times$ \textsc{Group Type} ($F(1, 114) = 5.35, p = 0.023$). Based on these results, there is an indication that dark patterns affected our non-ADHD participants' choice to choose the premium status more than our ADHD group.


\subsection{Avoidance in Task 2---Personal Data Entered}

\begin{figure*}[t!]
    \centering
    \includegraphics[width=1\textwidth]{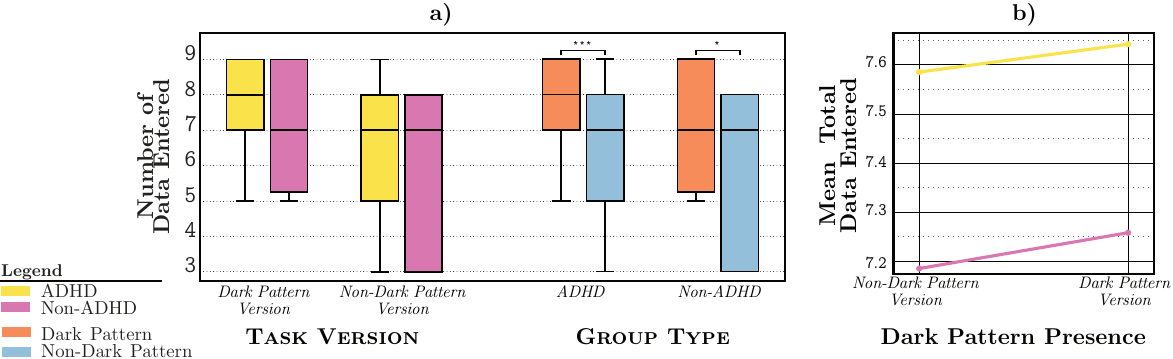}
    \caption{This figure shows the results of the entered data between \textsc{Task Version} and \textsc{Group Type} of Task 2. In a), the two box plots show how much data the groups entered per \textsc{Task Version} with significant indicators between the versions. In b), means show the interaction between the factors textsc{Task Versions} and \textsc{Group Type}, indicating a higher mean for ADHD individuals.}
    \Description[Two plots visualising the entered data in Task 2.]{This figure shows the results of the entered data between Task Version and Group Type of Task 2. In a), the two box plots show how much data the groups entered per Task Version with significant indicators between the versions. In b), means show the interaction between the factors Task Versions and Group Type, indicating a higher mean for ADHD individuals.}
    \label{fig:results-entered-data}
\end{figure*}


While asking to leave a comment on available content, the second task required participants to first enter certain data to enable the commenting functionality, for which they could draw from the provided persona (see Appendix~\ref{app:survey-material}). In either version of this task, participants could enter nine data points -- three of which were mandatory, including first and family name as well as date of birth in line with common SNSs. The remaining six data fields were optional and gave us the opportunity to investigate how the presence of dark patterns influences participants in providing more sensitive data\footnote{While participants were asked to use the scenario and persona data when entering anything, we did not store any data other than boolean values of the data fields, indicating whether one contains data or is empty.}. The box plots in Figure~\ref{fig:results-entered-data} a) visualise the differences between our \textsc{Task Version} and \textsc{Group Type}, indicating that the ADHD group entered more data overall, while the dark pattern version influenced how much data was generally entered across the two groups. Figure~\ref{fig:results-entered-data} b) underlines these results by showing the interaction between \textsc{Task Version} $\times$ \textsc{Group Type}, suggesting generally higher vulnerability among ADHD participants in this regard. 
Importantly, we removed outliers from this analysis, which resulted in 16 rows being excluded. The results of an ART-ANOVA show a significant interaction effect of \textsc{Task Version} $\times$ \textsc{Group Type} on the amount of data entered, $F(1, 114)=10.78$, $p<0.01$). Note that as both factors are binary, a post-hoc procedure is not required.
In sum, these findings imply that our participants with ADHD were affected more by the dark patterns coercing them to add more data, even on non-required fields, than the non-ADHD group.

\subsection{Avoidance of Check-Boxes}
\input{tables/checkboxes-table}
A common but deceptive design strategy in SNSs is so-called \textit{Privacy Zuckering}~\cite{bosch_2016_privacy,brignull_2022_wayback_types} -- when the service providers trick users into sharing more data than necessary or desired. This specific dark pattern is especially problematic when coupled with \textit{Bad Defaults}~\cite{bosch_2016_privacy} -- when optional choices to share information, like personal data, are pre-selected by the service provider and have to be actively turned off. In our study, we hid three such checkboxes, indicating an agreement to (1) receive E-mails from third parties, (2) give access to and upload personal contacts, and (3) grant access to personal usage data\footnote{As with other implemented dark patterns between the tasks, we did not store any data other than boolean values of the checkboxes (on/off).}. 
Table~\ref{tab:checkbox_data}, offering a complete overview of our participants' choices, suggests certain differences between \textsc{Task Versions} but few for \textsc{Group Type}, supporting previous results. 

\paragraph{Uploading Contact Data Checkbox}
The results of an ART-ANOVA test testing the effect of \textsc{Task Version} $\times$ \textsc{Group Type} on selecting the checkbox show a significant interaction between \textsc{Task Version} $\times$ \textsc{Group Type} ($F(1, 114) = 4.70, p = 0.032$), suggesting that although dark patterns affected all participants to upload contacts, participants of either group experienced these dark patterns differently.

\paragraph{Personal Data Checkbox}
Unlike the first checkbox, an ART-ANOVA did not indicate any significant main effects or interactions when participants had the opportunity to choose between sharing personal data or not. Interestingly, neither \textsc{Task Version} nor \textsc{Group Type} seemed to affect our participants' choices in this regard.

\paragraph{Receiving E-Mail Updates Checkbox}
For the final checkbox, the ART-ANOVA result shows a significant main effect of \textsc{Task Version} ($F(1, 114) = 19.95, p < 0.001$) on our participants' choice to receive updates via E-Mail. However, there was no interaction effect between \textsc{Task Version} $\times$ \textsc{Group Type}, which offers no conclusions in terms of comparing the ADHD group and the non-ADHD group.

\section{Discussion}
In this work, we present the results of a multi-factorial study investigating differences between users with and without ADHD in terms of recognising dark patterns and avoiding them.
We thereby focus on SNSs, as they present ubiquitous media that have been assigned both positive (e.g. \cite{eagle2023you, lin_why_2011}) and negative traits (e.g.~\cite{sina_social_2022, mildner_ethical_2021}) affecting their users' well-being. While different behavioural factors have been well-understood, mainly in social science and public health contexts (e.g. \cite{sina_social_2022, lin_why_2011,brandtzaeg_typology_2011}), the methods available to us within the HCI domain help us better understand which design elements contribute to harmful user interactions. 
In the past, different studies have demonstrated that although people are generally able to recognise dark patterns~\cite{bongard-blanchy_i_2021, maier_dark_2020,di_geronimo_ui_2020}, including in the context of SNSs~\cite{mildner_defending_2023, schaffner_understanding_2022}, the often small effects imply that many would still fall victim to the nefarious effects of dark patterns. And while research has gained a better understanding of the perception of dark patterns, few studies considered vulnerable and specific populations~\cite{sanchez_chamorro_my_2024}, with calls made to address this gap~\cite{gray_mobilizing_2024}. To this end, our study is, to our knowledge, the first that explored dark patterns as design elements in an interactive context while focusing on a specific user group, i.e. individuals with ADHD. Thus, we answered previous calls, motivated by recent work in this line that describes dark patterns exploiting users' attention and engaging them in undesired interactions~\cite{monge_roffarello_defining_2023,mildner_about_2023}. Here, we answer our two research questions through the insights gained from our work before we discuss underlying implications. 

\subsection{Recognising Dark Patterns -- Answering RQ1}
The results of the first study show that, while ADHD individuals are vulnerable to dark patterns, this happens no more so than for non-ADHD individuals. Overall, both participant groups showed similar abilities to recognise dark patterns based on Mathur et al.’s~\cite{mathur_what_2021} characteristics, as indicated by our results between items I1-I6, not showing any significant differences between our \textsc{Group Types}. 
Interestingly, the effects of dark patterns on our groups' choices between \textsc{Task Variants} differed. While we noticed an interaction effect between \textsc{Group Version} $\times$ \textsc{Group Type} in Task 1, no such effect occurred for Task 2. As the order of tasks was fixed, these results could hint at a potential order bias or learning effect. However, we varied the types of dark patterns across the tasks, which might suggest that dark patterns in Task 1 were more persuasive than those in Task 2. Furthermore, the results of our NASA-TLX suggest that people with ADHD experienced less cognitive load across several subscales, including effort, frustration, and mental demand. These results suggest an effect on different perceptions of dark patterns between our groups that may affect well-being through differently experienced stress levels.

Furthermore, research on neurodivergent and ADHD individuals has predominantly been deficit-focused~\cite{spiel2019agency, stefanidi2021children, spiel2022adhd}, meaning that while studies explored potential disadvantages of these populations, less work has been conducted focusing on other aspects. 
To this end, our results align with previous calls for HCI research and design to not focus merely on symptoms, trying to ``fix'' a condition~\cite{spiel2021purpose}, but rather supporting neurodivergent interests, including of ADHD individuals, adopting a strengths-based approach~\cite{spiel2021purpose, spiel2019agency, spiel2022adhd, stefanidi2022designing, stefanidi2023children}. 

With regard to dark pattern research, our study provides empirical insights that mirror previous findings of comparable experiments~\cite{mildner_about_2023, bongard-blanchy_i_2021} but did not consider any specific population. Importantly, we see a difference between being able to differentiate between tasks that contain dark patterns or not and sufficiently recognising dark patterns. As with these prior studies, our participants' ratings of dark pattern characteristics (items I1-I6) were relatively neutral, suggesting no reliability to recognise them. We thus add to existing calls for better protection from dark pattern harms either through responsible design or through regulatory measures.

\subsection{Avoiding Dark Patterns -- Answering RQ2}
Knowing about our participants' general ability to recognise dark patterns in our SNSs-based experiment, the question remains how effective dark patterns were in coercing our participants' choices. Unlike the rather similar responses between \textsc{Group Types} in the recognition of dark patterns, our results surfaced significant differences in their ability to avoid them. Drawing from an existing ontology~\cite{gray_building_2024} and related work~\cite{mildner_about_2023,monge_roffarello_defining_2023}, our study setup contained several dark patterns that participants faced and could fall victim to. These included \textit{Interface Interference}, like \textit{Bad Defaults} or \textit{Forced Actions} such as \textit{Attention Capture}.

During the account creation in Task 1, we noticed a strong interaction effect indicating that the presence of dark patterns generally influenced our participants' choices. Moreover, the analysis showed that non-ADHD participants were more likely to be coerced in their decision to choose a premium version than ADHD participants, suggesting that ADHD individuals might be less susceptible to dark patterns during similar tasks. 
We speculate that non-ADHD individuals may be more affected by subtle coercive elements in dark patterns because they tend to engage more predictably with standard cues and defaults in decision-making, potentially due to their consistent attention regulation. In contrast, ADHD participants, whose attention and impulsivity may vary significantly~\cite{gregg2000definition, american2013diagnostic}, might not respond as predictably to the same cues, potentially diminishing the intended effect of these dark patterns. 
This might also link to strengths associated with ADHD, such as cognitive flexibility~\cite{schippers2024associations}.
In any case, we encourage future research to determine the exact reasons behind this different perception of certain dark patterns. 

Regardless of their group, however, we observed a generally strong effect of dark patterns on all participants' choices, which stretched through different comparisons, including data entered and active checkboxes, strengthening the implications of dark patterns to manipulate choices. When asked to comment on the presented content as part of Task 2, participants first had to add personal data to their profiles. Although the profile page contains nine fields participants could enter data into, only some were mandatory to proceed, depending on the \textsc{Task Version}. Regardless of whether the task contained dark patterns or not, however, we noticed a difference between the \textsc{Group Types}. The dark pattern version of the task included \textit{Interface Interference}, mainly in the form of \textit{Emotional and Sensory Manipulation}, as well as \textit{Forced Action} and \textit{Social Engineering} patterns (see Figure~\ref{fig:example-t2}) to coerce participants to add more data. These patterns manifested, for instance, through the display of a completion indicator around the profile's picture while a smiley face changed its frown to a smile, the more data participants added.

Based on this task, we noticed that the ADHD group entered more data overall, suggesting a stronger vulnerability among participants of this group to disclose personal data when not needed. In part, these differences could be linked to a tendency among ADHD individuals to hyperfocus or hyper-fixate on specific tasks~\cite{eagle2023you}.
In particular, while an intense focus on a specific task or content, especially if it is personally stimulating or rewarding, is not a diagnostic criterion for ADHD, it’s a widely reported behaviour among ADHD individuals~\cite{huang2022snapshot}.
Despite this being also an associated strength of ADHD~\cite{schippers2024associations, hupfeld2019living}, this focus might also lead individuals to overlook details or consequences, like disclosing more personal data than necessary, due to a narrowed attention scope. 
In our case, we speculate that this occurred due to the specific dark pattern strategy used, i.e. the reward of a happier smiley and acquiring more completion points when providing more personal data.
Following this direction, our study demonstrates a particular vulnerability that urges attention to protect ADHD individuals' privacy better. This can be addressed through responsible design that gives users more autonomy to decide which data to disclose. Unfortunately, incentives driven by surveillance capitalism~\cite{zuboff_surveillance_2023} and the understood inability among practitioners to act in line with their values and beliefs~\cite{gray_ethical_2019, chivukula_identity_2021} suggests that additional safety measures are needed. To this end, policymakers and regulators should level the playing field to the degree that SNSs do not exploit their users.

\subsection{Further Implications}
Throughout our study, we noticed various significant effects of dark patterns on our participants' choices. While various regulations, including the DSA~\cite{eu_dsa_2022}, require service providers to give their users autonomy to make informed decisions, our results strongly indicate that the presence of dark patterns plays antagonist roles, infringing existing regulations. This incision into people's choice architecture, manipulating their decisions, is highly problematic regardless of population, especially for vulnerable and specific groups.

The growing adoption of SNSs has led to increased representation of neurodivergent community members with ADHD in these platforms~\cite{eagle2023you}. Given that SNSs can serve as a way for ADHD individuals to support each other and even share their symptoms and encourage others to seek diagnosis, the influence and importance of SNSs and online spaces becomes particularly important. Given that our study showed how ADHD individuals recognise dark patterns to a comparable degree to non-ADHD individuals, with either showing vulnerability to their harms, as well as the tendency to avoid dark patterns differently depending on the context, our work underlines the necessity of ethical and responsible design and general increased protection, especially better tailored around vulnerable and specific groups of people.

\section{Future Work \& Limitations}
While we were careful to design a study that offers meaningful insights by conducting a pre-study and grounding our choices in related work, our research has several limitations that we disclose here. First and most importantly, we explore the effects of dark patterns on SNSs through a custom-developed web application. While it is increasingly difficult to conduct studies of this nature on real SNSs, because of technological and access restrictions as well as strong ethical concerns, our results should be taken tentatively and in consideration that participants interacted with a mock SNS. 

In the design of our tasks and the user interface, we drew inspiration from popular SNSs, largely from Facebook and TikTok, which have been the focus of various SNS-related studies in the past. While these two platforms reflect the evolution of popular SNSs throughout the past two decades, we did not consider various other platforms and their options for users to engage with them. To mitigate limiting effects in this regard, we designed two tasks that require similar steps in most SNSs. Nonetheless, it would be interesting to see how participants would interact with other, more specific tasks. We hope that our repository will support future work in this regard. 

To measure differences between our factors, we deployed several items, based on Mathur et al.'s~\cite{mathur_what_2021} dark pattern characteristics. Although we ground this approach in related work conducted by \citet{mildner_defending_2023} to make our results comparable, we acknowledge that the items are not a validated scale. To support our results, we also included the NASA-TLX, which represents a validated and often used tool within and outside HCI studies. However, with more quantitative research on dark patterns being conducted, a validated scale could not only strengthen similar studies but make studies more comparable across contexts.

During Task 2, where participants had to enter personal data before commenting on SNS content, the dark pattern and non-dark pattern versions differed in the sense that different numbers of items were mandatory. In the non-dark pattern version, participants had to enter three data points, while the dark pattern version required 5. This decision is motivated by the inclusion of \textit{Forced Action} dark patterns to better understand how it steers user choices. Seeing all participants obeyed the pattern's strategy, we respected and considered these differences in our analysis, showing the effectiveness of this pattern. Nonetheless, this difference poses a certain limitation to the comparability between the task versions. Studying dark patterns in comparative setups is challenging, especially when they manipulate the interface in ways that lead to contrasting user experiences. While we aimed to mitigate this challenge through our overall study design, and hope our mock SNS supports similar research, future studies could follow alternative methodologies to understand how dark patterns affect users, as more research is needed in this regard.

Another critical limitation is that participants of our ADHD group self-reported their diagnoses. In particular, Prolific, the service we used to recruit our participants, does not verify these reports. Moreover, ADHD is not a binary condition but occurs on a continuum~\cite{mclennan_understanding_2016}, which limits our results further. For several ethical concerns alone, a different study apparatus would be required, including medical or psychological supervision, to confirm an ADHD diagnosis and measure where on the ADHD continuum participants might be. However, it would be interesting to see how nuanced aspects within the ADHD continuum relate to our results. Furthermore, other neurodivergent characteristics would also be interesting to consider. Worried particularly about the effects attention-grabbing and engaging dark patterns have on ADHD, other types of dark patterns may affect specific and vulnerable populations differently. Our work is a first step toward understanding how specific populations are affected by dark patterns in the context of SNSs. Further, we did not assess our participants' digital or technological literacy. Although we asked them about the frequency with which they used SNSs at the time of conducting the experiment, which might indicate literacy to a certain degree, additional insights could help draw implications between our results and specific demographics. We hope that future work can build on this work and advance our understanding to protect everyone regardless of context and as their needs afford.

\section{Conclusion}
In this paper, we presented a multi-factorial study investigating the ability of individuals with and without ADHD to recognise and avoid dark patterns in the context of social networking sites (SNSs). Contributing to a growing body of research on dark patterns, our work demonstrates their effects on vulnerable populations, specifically ADHD individuals. Our results suggest that both ADHD and non-ADHD participants exhibit comparable abilities in recognising dark patterns in SNS contexts. While the general ability is reflected in significant differences between considered factors, we argue that ratings were relatively neutral, suggesting that the perception to differentiate may not be sufficient in shielding SNS users from dark patterns. This argument is underlined by our results regarding the avoidance of dark patterns, which demonstrate their strong persuasiveness to steer user choices. Overall, ADHD participants showed decreased cognitive load but a heightened vulnerability in certain tasks, particularly those involving the disclosure of personal data. However, they were less likely to be persuaded into buying a premium version on the mock SNS platform. This implies that the context in which dark patterns occur influences how individuals with ADHD are able to avoid them.

Through our work, we highlight the need for better protection of all SNS users, with a specific focus on ADHD populations, against deceptive and manipulative strategies and dark patterns. The significant interaction effects we observed between task variation and group types imply the subtle influence of design features on user choices, raising concerns about their autonomy. 
Moreover, the findings suggest that current regulations, like the Digital Services Act (DSA), while important, may not be sufficient in safeguarding vulnerable users from the manipulative tactics embedded in digital environments.
While previous studies have addressed dark patterns' general impact, our study contributes to the current discourse by exploring specific effects on vulnerable populations, such as individuals with ADHD.
This calls for more stringent enforcement and possibly the development of new regulations to ensure SNSs uphold ethical design practices that protect user privacy and agency.

\begin{acks}
This work was partially supported by the Leibniz ScienceCampus Bremen Digital Public Health, which is jointly funded by the Leibniz Association (W72/2022), the Federal State of Bremen, and the Leibniz Institute for Prevention Research and Epidemiology – BIPS.
This research was also partly funded by the German Research Foundation (DFG) under Germany's Excellence Strategy (EXC 2077, University of Bremen).
We acknowledge the use of generative AI tools as writing assistants in the writing of this text. Specifically, we used the tools Grammarly and ChatGPT 4, but only to enhance the readability of our text by assisting with grammatical structuring and spelling corrections. All ideas, test statistics, and content remain the authors' work.
\end{acks}

\bibliographystyle{ACM-Reference-Format}
\bibliography{references.bib}

\appendix
\input{Appendix}

\end{document}

%% file: tables/mathur_characteristics.tex
\begin{table}[t]
\resizebox{1\linewidth}{!}{
\renewcommand{\arraystretch}{1.4}
\begin{tabular}{p{0.05\linewidth}p{0.175\linewidth}p{0.675\linewidth}}
\toprule
\multicolumn{3}{c}{{\begin{tabular}[c]{@{}c@{}}\textbf{Dark Pattern Characteristics}\\ by \citet{mathur_dark_2019} \& \citet{mathur_what_2021}\end{tabular}}}                                \\ 

\midrule
\textbf{Item} & \textbf{Char.}   & \textbf{Adapted statements used in our survey}                                                                                                                          \\ \midrule
\cellcolor[gray]{0.95}I1    & \cellcolor[gray]{0.95}Asymmetric          & \cellcolor[gray]{0.95}I found certain design choices to present available options unequally. \\
I2 & Covert                 & I found the effects of certain design choices hidden from me.\\
\cellcolor[gray]{0.95}I3  & \cellcolor[gray]{0.95}Deceptive           & \cellcolor[gray]{0.95}I found some design choices to induce false beliefs by confusing, misleading, or keeping information from me.\\
I4 & Hides Information      & I found certain design choices to obscure or delay necessary
information. \\
\cellcolor[gray]{0.95}I5    &\cellcolor[gray]{0.95}Restrictive         & \cellcolor[gray]{0.95}I found certain design choices to restrict available options.  \\ 
I6 & Disparate Treatment       & I found certain design choices to create (dis)advantages for user
groups or treating them differently. \\
\midrule
\cellcolor[gray]{0.95}I7 &\cellcolor[gray]{0.95}Definition & \cellcolor[gray]{0.95}``[U]ser interface design choices that benefit an online service by coercing, steering, or deceiving users into making decisions that, if fully informed and capable of selecting alternatives, they might not make'' -- \citet[p. 2]{mathur_dark_2019}\\
\bottomrule
\end{tabular}
}
\caption{This table presents our adapted statements (I1-I6) for each of the six characteristics as well as a general dark pattern definition (I7) from \citet{mathur_dark_2019} and \citet{mathur_what_2021} used in our survey 
in a Likert-scale fashion from 1 (not at all) - 7 (extremely).}
\Description[Dark Pattern Characteristics]{This table presents the six dark pattern characteristics from Mathur et al. (2021) and the statements we asked our participants to evaluate using a Likert-scale from 1 (not at all) - 7 (extremely).}
\vspace{-0.4cm}
\label{tab:characteristics_suvey}
\end{table}

%% file: tables/task-permutation.tex
\begin{table*}[t!]
\centering
\begin{tabular}{llcccc}
\toprule
\textbf{Task 1}        & \textbf{Task 2}        & \textbf{ADHD} & \textbf{Non-ADHD} & \textbf{Total} & \textbf{Version Total}\\
\midrule
\cellcolor[gray]{0.95}Dark Pattern  & \cellcolor[gray]{0.95}Dark Pattern    & \cellcolor[gray]{0.95}18   & \cellcolor[gray]{0.95} 17     & \cellcolor[gray]{0.95}35 &  \multirow{2}{*}{68} \\   
Dark Pattern  & Non-Dark Pattern & 17   & 16     & 33 \\ \hline
\cellcolor[gray]{0.95}Non-Dark Pattern & \cellcolor[gray]{0.95}Dark Pattern    & \cellcolor[gray]{0.95}17   & \cellcolor[gray]{0.95}17     & \cellcolor[gray]{0.95}34 &  \multirow{2}{*}{67}\\ 
Non-Dark Pattern & Non-Dark Pattern & 16   & 17     & 33 & \\
\midrule
 & \textbf{Total}   & 68   & 67     & 135 &  \\
\bottomrule
\end{tabular}
\caption{Participant distribution between groups and tasks.}
\vspace{-0.4cm}
\label{tab:task-permutation}
\end{table*}

%% file: tables/anova-results-q1.tex
\begin{table}[t!]
\resizebox{1\linewidth}{!}{
    \centering
    \begin{tabular}{lcccc}
        \toprule
        &\multicolumn{4}{c}{\textbf{Factors}}\\\cmidrule{2-5}
        &\multicolumn{2}{c}{\textbf{\textsc{Task Version}}}&\multicolumn{2}{c}{\textbf{\textsc{Group Type}}}\\
         & $F(1, 114)$  & $p$ & $F(1, 114)$ & $p$ \\

        \midrule
        \textit{I1 -- Asymmetric (t1)} & 9.67 & \textbf{<.01} & 0.18 & 0.669 \\
        \textit{I1 -- Asymmetric (t2)} & 0.05 & 0.821 & 0.07 & 0.793 \\
        \midrule
        \textit{I2 -- Covert (t1)} & 15.89 &\textbf{ <.001} & 1.45 & 0.232 \\
        \textit{I2 -- Covert (t2)} & 0.29 & 0.592 & 3.29 & 0.073 \\
        \midrule
        \textit{I3 -- Deceptive (t1)} & 7.75 & \textbf{<.01} & 0.01 & 0.945 \\
        \textit{I3 -- Deceptive (t2)} & 1.02 & 0.314 & 0.21 & 0.650 \\
        \midrule
        \textit{I4 -- Disparate Treat. (t1)} & 9.07 & \textbf{<.01 }& 0.01 & 0.904 \\
        \textit{I4 -- Disparate Treat. (t2)} & 0.00 & 0.995 & 0.30 & 0.5878 \\
        \midrule
        \textit{I5 -- Hides Info. (t1)} & 17.52 & \textbf{<.001} & 0.22 & 0.639 \\
        \textit{I5 -- Hides Info. (t2)} & 0.03 & 0.863 & 0.37 & 0.5469 \\
        \midrule
        \textit{I6 -- Restrictive (t1)} & 27.00 &\textbf{<.001} & 0.21 & 0.649 \\
        \textit{I6 -- Restrictive (t2)} & 0.23 & 0.634 & 0.26 & 0.614 \\
        \midrule\midrule
        \textit{I7 -- Definition} & 10.80 &\textbf{ <.01} & 3.85 & 0.052 \\
        \bottomrule
    \end{tabular}
    }
    \caption{ART-ANOVA results for items I1-I7 (see Table~\ref{tab:characteristics_suvey}) and the factors \textsc{Task Version} and \textsc{Group Type}. Results are split between t1 (Task 1) and t2 (Task 2), with significant factors highlighted in \textbf{bold}. As we did not obtain any interaction effects between the factors for these items (p>.05), we do not report these measures here.}
    \vspace{-0.77cm}
    \label{tab:anova_i1-7}
\end{table}

%% file: tables/anova-nasatlx.tex
\begin{table}[t]
\resizebox{1\linewidth}{!}{
\centering
\begin{tabular}{lrcccc}
\toprule
\textbf{NASA TLX}  &    &&\\
\textbf{Subscale}  & \textbf{Factor}     & \textbf{F (1, 114)} & \textbf{$p$}\\
\midrule
\textbf{Effort}              & \textsc{Group Type}                                   & 7.97             & \textbf{0.005}         \\
                             & \textsc{Task Version}                         & 0.68             & 0.412            \\
                             & \textsc{Group Type} $\times$ \textsc{Task Version}             & 1.91             & 0.168            \\
\midrule
\textbf{Frustration}         & \textsc{Group Type}                                 & 5.03             & \textbf{0.026}          \\
                             & \textsc{Task Version}                        & 6.38             & \textbf{0.012}          \\
                             & \textsc{Group Type} $\times$ \textsc{Task Version}            & 0.02             & 0.879            \\
\midrule
\textbf{Mental}       & \textsc{Group Type}                                   & 1.65             & 0.201            \\
\textbf{Demand}                             & \textsc{Task Version}                          & 5.24             & \textbf{0.023   }         \\
                             & \textsc{Group Type} $\times$ \textsc{Task Version}           & 0.00             & 0.960            \\
\midrule
\textbf{Physical}     & \textsc{Group Type}                                 & 6.70             & \textbf{0.010  }          \\
\textbf{Demand}                             & \textsc{Task Version}                         & 1.48             & 0.226            \\
                             & \textsc{Group Type} $\times$ \textsc{Task Version}           & 0.00             & 0.974            \\
\midrule
\textbf{Performance}         & \textsc{Group Type}                                  & 0.12             & 0.725            \\
                             & \textsc{Task Version}                        & 1.65             & 0.200            \\
                             & \textsc{Group Type} $\times$ \textsc{Task Version}            & 0.93             & 0.337            \\
\midrule
\textbf{Temporal}     & \textsc{Group Type}                                  & 2.56             & 0.111            \\
\textbf{Demand}                             & \textsc{Task Version}                           & 2.76             & 0.098          \\
                             & \textsc{Group Type} $\times$ \textsc{Task Version}           & 0.06             & 0.809            \\
\bottomrule
\end{tabular}
}
\caption{ART-ANOVA results for the NASA-TLX with significant effects highlighted in bold.}
\vspace{-0.7cm}
\label{tab:anova_nasa-tlx}
\end{table}

%% file: tables/checkboxes-table.tex
\begin{table*}[t]
\centering
\begin{tabular}{@{}llcccccccc@{}}
\toprule
\textbf{\textsc{Task Version}}          & \textbf{\textsc{Group Type}}    & \multicolumn{2}{c}{\textbf{Contacts}}            & \multicolumn{2}{c}{\textbf{Track Data}}           & \multicolumn{2}{c}{\textbf{Updates}}            \\
                 &          &$\boxtimes$ & $\square$  & $\boxtimes$ & $\square$ & $\boxtimes$ & $\square$ \\
\midrule
\rowcolor[gray]{0.95}Dark Pattern     & ADHD     & 34      & 1        & 15      & 20       & 33      & 2         \\
Dark Pattern     & Non-ADHD & 33      & 0         &  19      & 15        & 32      & 1         \\ \midrule
\rowcolor[gray]{0.95}Non-Dark Pattern & ADHD     & 15      & 18       & 10      & 23        & 19      & 14         \\
Non-Dark Pattern & Non-ADHD & 14      & 20          & 10      & 23       & 18      & 16        \\
\bottomrule
\end{tabular}%
\caption{This table offers an overview of the checkbox selections between \textsc{Group Type} and \textsc{Task Version}. Here, the $\boxtimes$ symbol indicates checked, whereas the $\square$ represents unchecked options. The Contacts column contains the results for the `upload contacts checkbox', the Track Data for `allow personal data tracking', and Updates includes `receiving updates from our selected partners.'}
\label{tab:checkbox_data}
\end{table*}

%% file: Appendix.tex
\section{Survey Material}\label{app:survey-material}
To avoid participants entering their personal information during the web-study, we presented them with a scenario and gave them a persona to use during the study. This further allowed us to mitigate choices based on individual preferences and circumstances. As both scenario and persona were displayed on a desktop device, they stayed available during the tasks, which were completed on a second, mobile device. 

\subsection{Scenario}
Before doing the two tasks on the web-application, we presented our participants with the following brief scenario:

\vspace{11pt}
\textit{Please imagine that you are at home after a long day and decided to get back in touch with your family and friends. However, everyone is using a new social media application. So before you can re-connect with them, you need to create a new account (first task).}

\vspace{11pt}
\textit{After you successfully create an account and enter the social media app, you want to leave a nice comment on a post that you see (second task).}

\subsection{Persona}
All participants received the same persona:

\input{tables/persona}

\vspace{10cm}

%% file: tables/persona.tex
\begin{table}[h!]
\centering
\begin{tabular}{ll}
\toprule
\multicolumn{2}{c}{\textbf{Persona}} \\ 
\midrule
\textbf{First name:} & Charlie \\
\textbf{Last name:} & Harper \\
\textbf{Gender:} & Female \\
\textbf{Birthday:} & March 20, 1985 \\
\textbf{Phone:} & 0555 1234567 \\
\textbf{City:} & Rivertown \\
\textbf{Job title:} & Creative Director \\
\textbf{Email:} & charlie.harper@awesomemail.com \\
\textbf{Password:} & Sparkle123! \\
\bottomrule
\end{tabular}
\caption{This table presents the persona-data participants received during the survey to be used during the web-study.}
\end{table}

%% file: main.bbl

\begin{thebibliography}{90}


\ifx \showCODEN    \undefined \def \showCODEN     #1{\unskip}     \fi
\ifx \showISBNx    \undefined \def \showISBNx     #1{\unskip}     \fi
\ifx \showISBNxiii \undefined \def \showISBNxiii  #1{\unskip}     \fi
\ifx \showISSN     \undefined \def \showISSN      #1{\unskip}     \fi
\ifx \showLCCN     \undefined \def \showLCCN      #1{\unskip}     \fi
\ifx \shownote     \undefined \def \shownote      #1{#1}          \fi
\ifx \showarticletitle \undefined \def \showarticletitle #1{#1}   \fi
\ifx \showURL      \undefined \def \showURL       {\relax}        \fi
\providecommand\bibfield[2]{#2}
\providecommand\bibinfo[2]{#2}
\providecommand\natexlab[1]{#1}
\providecommand\showeprint[2][]{arXiv:#2}

\bibitem[Ahn and Shin(2013)]%
        {ahn2013social}
\bibfield{author}{\bibinfo{person}{Dohyun Ahn} {and} \bibinfo{person}{Dong-Hee Shin}.} \bibinfo{year}{2013}\natexlab{}.
\newblock \showarticletitle{Is the social use of media for seeking connectedness or for avoiding social isolation? Mechanisms underlying media use and subjective well-being}.
\newblock \bibinfo{journal}{\emph{Computers in Human Behavior}} \bibinfo{volume}{29}, \bibinfo{number}{6} (\bibinfo{year}{2013}), \bibinfo{pages}{2453--2462}.
\newblock


\bibitem[American Psychiatric~Association et~al\mbox{.}(2013)]%
        {american2013diagnostic}
\bibfield{author}{\bibinfo{person}{DS American Psychiatric~Association}, \bibinfo{person}{American~Psychiatric Association}, {et~al\mbox{.}}} \bibinfo{year}{2013}\natexlab{}.
\newblock \bibinfo{booktitle}{\emph{Diagnostic and statistical manual of mental disorders: DSM-5}}. Vol.~\bibinfo{volume}{5}.
\newblock \bibinfo{publisher}{American psychiatric association Washington, DC}.
\newblock


\bibitem[{Association for Computing Machinery}(2023)]%
        {acm_words_2023}
\bibfield{author}{\bibinfo{person}{{Association for Computing Machinery}}.} \bibinfo{year}{2023}\natexlab{}.
\newblock \bibinfo{booktitle}{\emph{Words matter: {Alternatives} for charged terminology in the computing profession}}.
\newblock
\urldef\tempurl%
\url{https://www.acm.org/diversity-inclusion/words-matter}
\showURL{%
\tempurl}
\newblock
\shownote{Visited on 2023-07-31}.


\bibitem[Aydin et~al\mbox{.}(2024)]%
        {aydin_trait-level_2024}
\bibfield{author}{\bibinfo{person}{Tuba Aydin}, \bibinfo{person}{Benjamin~A. Parris}, \bibinfo{person}{Gizem Arabaci}, \bibinfo{person}{Marina Kilintari}, {and} \bibinfo{person}{Jacqui Taylor}.} \bibinfo{year}{2024}\natexlab{}.
\newblock \showarticletitle{Trait-level non-clinical {ADHD} symptoms in a community sample and their association with technology addictions}.
\newblock  \bibinfo{volume}{43}, \bibinfo{number}{12} (\bibinfo{year}{2024}), \bibinfo{pages}{10682--10692}.
\newblock
\showISSN{1936-4733}
\href{https://doi.org/10.1007/s12144-023-05203-x}{doi:\nolinkurl{10.1007/s12144-023-05203-x}}


\bibitem[Bongard-Blanchy et~al\mbox{.}(2021)]%
        {bongard-blanchy_i_2021}
\bibfield{author}{\bibinfo{person}{Kerstin Bongard-Blanchy}, \bibinfo{person}{Arianna Rossi}, \bibinfo{person}{Salvador Rivas}, \bibinfo{person}{Sophie Doublet}, \bibinfo{person}{Vincent Koenig}, {and} \bibinfo{person}{Gabriele Lenzini}.} \bibinfo{year}{2021}\natexlab{}.
\newblock \showarticletitle{I am {Definitely} {Manipulated}, {Even} {When} {I} am {Aware} of it. {It} s {Ridiculous}! -- {Dark} {Patterns} from the {End}-{User} {Perspective}}.
\newblock \bibinfo{journal}{\emph{Designing Interactive Systems Conference 2021}} (\bibinfo{date}{June} \bibinfo{year}{2021}), \bibinfo{pages}{763--776}.
\newblock
\href{https://doi.org/10.1145/3461778.3462086}{doi:\nolinkurl{10.1145/3461778.3462086}}


\bibitem[B{\"o}sch et~al\mbox{.}(2016)]%
        {bosch_2016_privacy}
\bibfield{author}{\bibinfo{person}{Christoph B{\"o}sch}, \bibinfo{person}{Benjamin Erb}, \bibinfo{person}{Frank Kargl}, \bibinfo{person}{Henning Kopp}, {and} \bibinfo{person}{Stefan Pfattheicher}.} \bibinfo{year}{2016}\natexlab{}.
\newblock \showarticletitle{Tales from the Dark Side: Privacy Dark Strategies and Privacy Dark Patterns.}
\newblock \bibinfo{journal}{\emph{Proc. Priv. Enhancing Technol.}} \bibinfo{volume}{2016}, \bibinfo{number}{4} (\bibinfo{year}{2016}), \bibinfo{pages}{237--254}.
\newblock


\bibitem[Brandtzæg and Heim(2011)]%
        {brandtzaeg_typology_2011}
\bibfield{author}{\bibinfo{person}{Petter~Bae Brandtzæg} {and} \bibinfo{person}{Jan Heim}.} \bibinfo{year}{2011}\natexlab{}.
\newblock \showarticletitle{A typology of social networking sites users}.
\newblock  \bibinfo{volume}{7}, \bibinfo{number}{1} (\bibinfo{year}{2011}), \bibinfo{pages}{28--51}.
\newblock
\href{https://doi.org/10.1504/IJWBC.2011.038124}{doi:\nolinkurl{10.1504/IJWBC.2011.038124}}


\bibitem[Brignull(2010)]%
        {brignull_2022_wayback_types}
\bibfield{author}{\bibinfo{person}{Harry Brignull}.} \bibinfo{year}{2010}\natexlab{}.
\newblock \bibinfo{booktitle}{\emph{Deceptive design - Types of deceptive design}}.
\newblock
\urldef\tempurl%
\url{http://web.archive.org/web/20220525230009/https://www.deceptive.design/types}
\showURL{%
\tempurl}
\newblock
\shownote{Visited on 2023-07-27}.


\bibitem[Charabin et~al\mbox{.}(2023)]%
        {charabin2023m}
\bibfield{author}{\bibinfo{person}{Emma Charabin}, \bibinfo{person}{Emma~A Climie}, \bibinfo{person}{Courtney Miller}, \bibinfo{person}{Kristina Jelinkova}, {and} \bibinfo{person}{Jessica Wilkins}.} \bibinfo{year}{2023}\natexlab{}.
\newblock \showarticletitle{“I’m doing okay”: Strengths and resilience of children with and without ADHD}.
\newblock \bibinfo{journal}{\emph{Journal of Attention Disorders}} \bibinfo{volume}{27}, \bibinfo{number}{9} (\bibinfo{year}{2023}), \bibinfo{pages}{1009--1019}.
\newblock


\bibitem[Chivukula et~al\mbox{.}(2021)]%
        {chivukula_identity_2021}
\bibfield{author}{\bibinfo{person}{Shruthi~Sai Chivukula}, \bibinfo{person}{Aiza Hasib}, \bibinfo{person}{Ziqing Li}, \bibinfo{person}{Jingle Chen}, {and} \bibinfo{person}{Colin~M. Gray}.} \bibinfo{year}{2021}\natexlab{}.
\newblock \showarticletitle{Identity {Claims} that {Underlie} {Ethical} {Awareness} and {Action}}. In \bibinfo{booktitle}{\emph{Proceedings of the 2021 {CHI} {Conference} on {Human} {Factors} in {Computing} {Systems}}} \emph{(\bibinfo{series}{{CHI} '21})}. \bibinfo{publisher}{Association for Computing Machinery}, \bibinfo{address}{New York, NY, USA}, \bibinfo{pages}{1--13}.
\newblock
\showISBNx{978-1-4503-8096-6}
\href{https://doi.org/10.1145/3411764.3445375}{doi:\nolinkurl{10.1145/3411764.3445375}}


\bibitem[Cho et~al\mbox{.}(2002)]%
        {cho2002attention}
\bibfield{author}{\bibinfo{person}{Baek~Hwan Cho}, \bibinfo{person}{Jong-Min Lee}, \bibinfo{person}{JH Ku}, \bibinfo{person}{Dong~Pyo Jang}, \bibinfo{person}{JS Kim}, \bibinfo{person}{In-Young Kim}, \bibinfo{person}{Jang-Han Lee}, {and} \bibinfo{person}{Sun~I Kim}.} \bibinfo{year}{2002}\natexlab{}.
\newblock \showarticletitle{Attention enhancement system using virtual reality and EEG biofeedback}. In \bibinfo{booktitle}{\emph{Proceedings IEEE Virtual Reality 2002}}. IEEE, \bibinfo{pages}{156--163}.
\newblock


\bibitem[Cibrian et~al\mbox{.}(2020)]%
        {cibrian2020research}
\bibfield{author}{\bibinfo{person}{Franceli~L Cibrian}, \bibinfo{person}{Gillian~R Hayes}, {and} \bibinfo{person}{Kimberley~D Lakes}.} \bibinfo{year}{2020}\natexlab{}.
\newblock \showarticletitle{Research Advances in ADHD and Technology}.
\newblock \bibinfo{journal}{\emph{Synthesis Lectures on Assistive, Rehabilitative. and Health-Preserving Technologies}} \bibinfo{volume}{9}, \bibinfo{number}{3} (\bibinfo{year}{2020}), \bibinfo{pages}{i--156}.
\newblock
\href{https://doi.org/10.2200/S01061ED1V01Y202011ARH015}{doi:\nolinkurl{10.2200/S01061ED1V01Y202011ARH015}}


\bibitem[Cibrian et~al\mbox{.}(2022)]%
        {cibrian2022potential}
\bibfield{author}{\bibinfo{person}{Franceli~L Cibrian}, \bibinfo{person}{Kimberley~D Lakes}, \bibinfo{person}{Sabrina~EB Schuck}, {and} \bibinfo{person}{Gillian~R Hayes}.} \bibinfo{year}{2022}\natexlab{}.
\newblock \showarticletitle{The potential for emerging technologies to support self-regulation in children with ADHD: A literature review}.
\newblock \bibinfo{journal}{\emph{International Journal of Child-Computer Interaction}}  \bibinfo{volume}{31} (\bibinfo{year}{2022}), \bibinfo{pages}{100421}.
\newblock
\href{https://doi.org/10.1016/j.ijcci.2021.100421}{doi:\nolinkurl{10.1016/j.ijcci.2021.100421}}


\bibitem[Commission(2024)]%
        {european_commission_commission_nodate}
\bibfield{author}{\bibinfo{person}{European Commission}.} \bibinfo{year}{2024}\natexlab{}.
\newblock \bibinfo{booktitle}{\emph{Commission opens formal proceedings against X under the {DSA}}}.
\newblock
\urldef\tempurl%
\url{https://ec.europa.eu/commission/presscorner/detail/en/IP_23_6709}
\showURL{%
\tempurl}
\newblock
\shownote{Accessed: 09/12/2024}.


\bibitem[Commission(2022)]%
        {FTC2022}
\bibfield{author}{\bibinfo{person}{Federal~Trade Commission}.} \bibinfo{year}{2022}\natexlab{}.
\newblock \bibinfo{booktitle}{\emph{Bringing Dark Patterns to Light Staff Report}}.
\newblock \bibinfo{type}{{T}echnical {R}eport}. \bibinfo{institution}{Federal Trade Commission}.
\newblock
\urldef\tempurl%
\url{https://www.ftc.gov/system/files/ftc_gov/pdf/P214800\%20Dark\%20Patterns\%20Report\%209.14.2022\%20-\%20FINAL.pdf}
\showURL{%
\tempurl}


\bibitem[Dalton(2013)]%
        {dalton2013neurodiversity}
\bibfield{author}{\bibinfo{person}{Nicholas~Sheep Dalton}.} \bibinfo{year}{2013}\natexlab{}.
\newblock \showarticletitle{Neurodiversity \& hci}.
\newblock In \bibinfo{booktitle}{\emph{CHI'13 Extended abstracts on human factors in computing systems}}. \bibinfo{pages}{2295--2304}.
\newblock


\bibitem[Di~Geronimo et~al\mbox{.}(2020)]%
        {di_geronimo_ui_2020}
\bibfield{author}{\bibinfo{person}{Linda Di~Geronimo}, \bibinfo{person}{Larissa Braz}, \bibinfo{person}{Enrico Fregnan}, \bibinfo{person}{Fabio Palomba}, {and} \bibinfo{person}{Alberto Bacchelli}.} \bibinfo{year}{2020}\natexlab{}.
\newblock \showarticletitle{{UI} {Dark} {Patterns} and {Where} to {Find} {Them}}. In \bibinfo{booktitle}{\emph{Proceedings of the 2020 {CHI} {Conference} on {Human} {Factors} in {Computing} {Systems}}}. \bibinfo{publisher}{ACM}, \bibinfo{address}{New York, NY, USA}, \bibinfo{pages}{1--14}.
\newblock
\showISBNx{978-1-4503-6708-0}
\href{https://doi.org/10.1145/3313831.3376600}{doi:\nolinkurl{10.1145/3313831.3376600}}


\bibitem[Eagle and Ringland(2023)]%
        {eagle2023you}
\bibfield{author}{\bibinfo{person}{Tessa Eagle} {and} \bibinfo{person}{Kathryn~E Ringland}.} \bibinfo{year}{2023}\natexlab{}.
\newblock \showarticletitle{“You Can't Possibly Have ADHD”: Exploring Validation and Tensions around Diagnosis within Unbounded ADHD Social Media Communities}. In \bibinfo{booktitle}{\emph{Proceedings of the 25th International ACM SIGACCESS Conference on Computers and Accessibility}}. \bibinfo{pages}{1--17}.
\newblock


\bibitem[{European Parliament}(2022)]%
        {eu_dsa_2022}
\bibfield{author}{\bibinfo{person}{{European Parliament}}.} \bibinfo{year}{2022}\natexlab{}.
\newblock \bibinfo{title}{Digital Servicess Act***I. European Parliament [A9-0356/2021]}.
\newblock
\urldef\tempurl%
\url{https://www.europarl.europa.eu/doceo/document/TA-9-2022-0014_EN.html}
\showURL{%
\tempurl}
\newblock
\shownote{Visited on 2024-01-26}.


\bibitem[Faraone et~al\mbox{.}(2015)]%
        {faraone2015attention}
\bibfield{author}{\bibinfo{person}{SV Faraone}, \bibinfo{person}{P Asherson}, \bibinfo{person}{T Banaschewski}, \bibinfo{person}{J Biederman}, \bibinfo{person}{JK Buitelaar}, \bibinfo{person}{JA Ramos-Quiroga}, {and} \bibinfo{person}{B Franke}.} \bibinfo{year}{2015}\natexlab{}.
\newblock \bibinfo{title}{Attention-deficit/hyperactivity disorder. Nature Rev. Dis. Primers 15020}.
\newblock


\bibitem[Faul et~al\mbox{.}(2009)]%
        {faul2009gpower}
\bibfield{author}{\bibinfo{person}{Franz Faul}, \bibinfo{person}{Edgar Erdfelder}, \bibinfo{person}{Axel Buchner}, {and} \bibinfo{person}{Alexander-Georg Lang}.} \bibinfo{year}{2009}\natexlab{}.
\newblock \bibinfo{booktitle}{\emph{G*Power (Version 3.1.9.7)}}.
\newblock
\urldef\tempurl%
\url{https://www.gpower.hhu.de}
\showURL{%
\tempurl}
\newblock
\shownote{Computer software}.


\bibitem[Friedman et~al\mbox{.}(2013)]%
        {friedman_value_2013}
\bibfield{author}{\bibinfo{person}{Batya Friedman}, \bibinfo{person}{Peter~H. Kahn}, \bibinfo{person}{Alan Borning}, {and} \bibinfo{person}{Alina Huldtgren}.} \bibinfo{year}{2013}\natexlab{}.
\newblock \showarticletitle{Value {Sensitive} {Design} and {Information} {Systems}}.
\newblock In \bibinfo{booktitle}{\emph{Early engagement and new technologies: {Opening} up the laboratory}}, \bibfield{editor}{\bibinfo{person}{Neelke Doorn}, \bibinfo{person}{Daan Schuurbiers}, \bibinfo{person}{Ibo van~de Poel}, {and} \bibinfo{person}{Michael~E. Gorman}} (Eds.). Vol.~\bibinfo{volume}{16}. \bibinfo{publisher}{Springer Netherlands}, \bibinfo{address}{Dordrecht}, \bibinfo{pages}{55--95}.
\newblock
\showISBNx{978-94-007-7843-6 978-94-007-7844-3}
\href{https://doi.org/10.1007/978-94-007-7844-3_4}{doi:\nolinkurl{10.1007/978-94-007-7844-3_4}}
\newblock
\shownote{Series Title: Philosophy of Engineering and Technology}.


\bibitem[Furman(2005)]%
        {furman2005attention}
\bibfield{author}{\bibinfo{person}{Lydia Furman}.} \bibinfo{year}{2005}\natexlab{}.
\newblock \showarticletitle{What is attention-deficit hyperactivity disorder (ADHD)?}
\newblock \bibinfo{journal}{\emph{Journal of child neurology}} \bibinfo{volume}{20}, \bibinfo{number}{12} (\bibinfo{year}{2005}), \bibinfo{pages}{994--1002}.
\newblock


\bibitem[Gray and Chivukula(2019)]%
        {gray_ethical_2019}
\bibfield{author}{\bibinfo{person}{Colin~M. Gray} {and} \bibinfo{person}{Shruthi~Sai Chivukula}.} \bibinfo{year}{2019}\natexlab{}.
\newblock \showarticletitle{Ethical {Mediation} in {UX} {Practice}}. In \bibinfo{booktitle}{\emph{Proceedings of the 2019 {CHI} {Conference} on {Human} {Factors} in {Computing} {Systems}}}. \bibinfo{publisher}{ACM}, \bibinfo{address}{Glasgow Scotland Uk}, \bibinfo{pages}{1--11}.
\newblock
\showISBNx{978-1-4503-5970-2}
\href{https://doi.org/10.1145/3290605.3300408}{doi:\nolinkurl{10.1145/3290605.3300408}}


\bibitem[Gray et~al\mbox{.}(2024a)]%
        {gray_mobilizing_2024}
\bibfield{author}{\bibinfo{person}{Colin~M. Gray}, \bibinfo{person}{Johanna~T. Gunawan}, \bibinfo{person}{Ren\'{e} Sch\"{a}fer}, \bibinfo{person}{Nataliia Bielova}, \bibinfo{person}{Lorena Sanchez~Chamorro}, \bibinfo{person}{Katie Seaborn}, \bibinfo{person}{Thomas Mildner}, {and} \bibinfo{person}{Hauke Sandhaus}.} \bibinfo{year}{2024}\natexlab{a}.
\newblock \showarticletitle{Mobilizing Research and Regulatory Action on Dark Patterns and Deceptive Design Practices}. In \bibinfo{booktitle}{\emph{Extended Abstracts of the 2024 CHI Conference on Human Factors in Computing Systems}} \emph{(\bibinfo{series}{CHI EA '24})}. \bibinfo{publisher}{Association for Computing Machinery}, \bibinfo{address}{New York, NY, USA}, Article \bibinfo{articleno}{482}, \bibinfo{numpages}{6}~pages.
\newblock
\showISBNx{9798400703317}
\href{https://doi.org/10.1145/3613905.3636310}{doi:\nolinkurl{10.1145/3613905.3636310}}


\bibitem[Gray et~al\mbox{.}(2018)]%
        {gray_dark_2018}
\bibfield{author}{\bibinfo{person}{Colin~M. Gray}, \bibinfo{person}{Yubo Kou}, \bibinfo{person}{Bryan Battles}, \bibinfo{person}{Joseph Hoggatt}, {and} \bibinfo{person}{Austin~L. Toombs}.} \bibinfo{year}{2018}\natexlab{}.
\newblock \showarticletitle{The Dark (Patterns) Side of UX Design}.
\newblock  (\bibinfo{year}{2018}), \bibinfo{pages}{1–14}.
\newblock
\showISBNx{9781450356206}
\href{https://doi.org/10.1145/3173574.3174108}{doi:\nolinkurl{10.1145/3173574.3174108}}


\bibitem[Gray et~al\mbox{.}(2023)]%
        {gray_temporal_2023}
\bibfield{author}{\bibinfo{person}{Colin~M. Gray}, \bibinfo{person}{Thomas Mildner}, {and} \bibinfo{person}{Nataliia Bielova}.} \bibinfo{year}{2023}\natexlab{}.
\newblock \showarticletitle{{Temporal Analysis of Dark Patterns: A Case Study of a User’s Odyssey to Conquer Prime Membership Cancellation through the ``Iliad Flow''}}. \bibinfo{publisher}{{arXiv}}.
\newblock
\showeprint[arxiv]{2309.09635 [cs]}
\urldef\tempurl%
\url{http://arxiv.org/abs/2309.09635}
\showURL{%
\tempurl}


\bibitem[Gray et~al\mbox{.}(2024b)]%
        {gray_building_2024}
\bibfield{author}{\bibinfo{person}{Colin~M. Gray}, \bibinfo{person}{Ike Obi}, \bibinfo{person}{Shruthi~Sai Chivukula}, \bibinfo{person}{Ziqing Li}, \bibinfo{person}{Thomas Carlock}, \bibinfo{person}{Matthew Will}, \bibinfo{person}{Anne~C Pivonka}, \bibinfo{person}{Janna Johns}, \bibinfo{person}{Brookley Rigsbee}, \bibinfo{person}{Ambika~R Menon}, {and} \bibinfo{person}{Aayushi Bharadwaj}.} \bibinfo{year}{2024}\natexlab{b}.
\newblock \showarticletitle{{Building an Ethics-Focused Action Plan: Roles, Process Moves, and Trajectories}}. In \bibinfo{booktitle}{\emph{roceedings of the {CHI} {Conference} on {Human} {Factors} in {Computing} {Systems} ({CHI} ’24)}} (2024). \bibinfo{publisher}{ACM, New York, NY, USA}, \bibinfo{address}{Honolulu, HI, USA}, \bibinfo{pages}{1--18}.
\newblock
\showISBNx{979-8-4007-0330-0/24/05}
\href{https://doi.org/10.1145/3613904.3642302}{doi:\nolinkurl{10.1145/3613904.3642302}}


\bibitem[Gray et~al\mbox{.}(2021)]%
        {gray_dark_2021}
\bibfield{author}{\bibinfo{person}{Colin~M. Gray}, \bibinfo{person}{Cristiana Santos}, \bibinfo{person}{Nataliia Bielova}, \bibinfo{person}{Michael Toth}, {and} \bibinfo{person}{Damian Clifford}.} \bibinfo{year}{2021}\natexlab{}.
\newblock \showarticletitle{Dark patterns and the legal requirements of consent banners: {An} interaction criticism perspective}, In \bibinfo{booktitle}{Proceedings of the 2021 CHI Conference on Human Factors in Computing Systems}.
\newblock \bibinfo{journal}{\emph{arXiv}}, Article \bibinfo{articleno}{172}, \bibinfo{numpages}{18}~pages.
\newblock
\showISBNx{9781450380966}
\showISSN{9781450380966}
\href{https://doi.org/10.1145/3411764.3445779}{doi:\nolinkurl{10.1145/3411764.3445779}}


\bibitem[Gray et~al\mbox{.}(2024c)]%
        {gray_ontology_2023}
\bibfield{author}{\bibinfo{person}{Colin~M. Gray}, \bibinfo{person}{Cristiana~Teixeira Santos}, \bibinfo{person}{Nataliia Bielova}, {and} \bibinfo{person}{Thomas Mildner}.} \bibinfo{year}{2024}\natexlab{c}.
\newblock \showarticletitle{An Ontology of Dark Patterns Knowledge: Foundations, Definitions, and a Pathway for Shared Knowledge-Building}. In \bibinfo{booktitle}{\emph{Proceedings of the CHI Conference on Human Factors in Computing Systems}} (Honolulu, HI, USA) \emph{(\bibinfo{series}{CHI '24})}. \bibinfo{publisher}{Association for Computing Machinery}, \bibinfo{address}{New York, NY, USA}, Article \bibinfo{articleno}{289}, \bibinfo{numpages}{22}~pages.
\newblock
\showISBNx{9798400703300}
\href{https://doi.org/10.1145/3613904.3642436}{doi:\nolinkurl{10.1145/3613904.3642436}}


\bibitem[Gregg and Scott(2000)]%
        {gregg2000definition}
\bibfield{author}{\bibinfo{person}{No{\"e}l Gregg} {and} \bibinfo{person}{Sally~S Scott}.} \bibinfo{year}{2000}\natexlab{}.
\newblock \showarticletitle{Definition and documentation: Theory, measurement, and the courts}.
\newblock \bibinfo{journal}{\emph{Journal of Learning Disabilities}} \bibinfo{volume}{33}, \bibinfo{number}{1} (\bibinfo{year}{2000}), \bibinfo{pages}{5--13}.
\newblock


\bibitem[Hansen and Jespersen(2013)]%
        {hansen_nudge_2013}
\bibfield{author}{\bibinfo{person}{Pelle~Guldborg Hansen} {and} \bibinfo{person}{Andreas~Maaløe Jespersen}.} \bibinfo{year}{2013}\natexlab{}.
\newblock \showarticletitle{Nudge and the {Manipulation} of {Choice}: {A} {Framework} for the {Responsible} {Use} of the {Nudge} {Approach} to {Behaviour} {Change} in {Public} {Policy}}.
\newblock \bibinfo{journal}{\emph{European Journal of Risk Regulation}} \bibinfo{volume}{4}, \bibinfo{number}{1} (\bibinfo{date}{March} \bibinfo{year}{2013}), \bibinfo{pages}{3--28}.
\newblock
\showISSN{1867-299X, 2190-8249}
\href{https://doi.org/10.1017/S1867299X00002762}{doi:\nolinkurl{10.1017/S1867299X00002762}}


\bibitem[Hart and Staveland(1988)]%
        {hart1988development}
\bibfield{author}{\bibinfo{person}{Sandra~G. Hart} {and} \bibinfo{person}{Lowell~E. Staveland}.} \bibinfo{year}{1988}\natexlab{}.
\newblock \bibinfo{booktitle}{\emph{Development of NASA-TLX (Task Load Index): Results of empirical and theoretical research}}.
\newblock \bibinfo{type}{{T}echnical {R}eport} NASA-TM-104174. \bibinfo{institution}{NASA Ames Research Center}, \bibinfo{address}{Moffett Field, CA}.
\newblock


\bibitem[Hoogman et~al\mbox{.}(2020)]%
        {hoogman2020creativity}
\bibfield{author}{\bibinfo{person}{Martine Hoogman}, \bibinfo{person}{Marije Stolte}, \bibinfo{person}{Matthijs Baas}, {and} \bibinfo{person}{Evelyn Kroesbergen}.} \bibinfo{year}{2020}\natexlab{}.
\newblock \showarticletitle{Creativity and ADHD: A review of behavioral studies, the effect of psychostimulants and neural underpinnings}.
\newblock \bibinfo{journal}{\emph{Neuroscience \& Biobehavioral Reviews}}  \bibinfo{volume}{119} (\bibinfo{year}{2020}), \bibinfo{pages}{66--85}.
\newblock


\bibitem[Huang(2022)]%
        {huang2022snapshot}
\bibfield{author}{\bibinfo{person}{Claire Huang}.} \bibinfo{year}{2022}\natexlab{}.
\newblock \showarticletitle{A snapshot into ADHD: The impact of hyperfixations and hyperfocus from adolescence to adulthood}.
\newblock \bibinfo{journal}{\emph{Journal of Student Research}} \bibinfo{volume}{11}, \bibinfo{number}{3} (\bibinfo{year}{2022}).
\newblock


\bibitem[Hupfeld et~al\mbox{.}(2019)]%
        {hupfeld2019living}
\bibfield{author}{\bibinfo{person}{Kathleen~E Hupfeld}, \bibinfo{person}{Tessa~R Abagis}, {and} \bibinfo{person}{Priti Shah}.} \bibinfo{year}{2019}\natexlab{}.
\newblock \showarticletitle{Living “in the zone”: hyperfocus in adult ADHD}.
\newblock \bibinfo{journal}{\emph{ADHD Attention Deficit and Hyperactivity Disorders}}  \bibinfo{volume}{11} (\bibinfo{year}{2019}), \bibinfo{pages}{191--208}.
\newblock


\bibitem[Leimstädtner et~al\mbox{.}(2023)]%
        {leimstadtner_investigating_2023}
\bibfield{author}{\bibinfo{person}{David Leimstädtner}, \bibinfo{person}{Peter Sörries}, {and} \bibinfo{person}{Claudia Müller-Birn}.} \bibinfo{year}{2023}\natexlab{}.
\newblock \showarticletitle{Investigating {Responsible} {Nudge} {Design} for {Informed} {Decision}-{Making} {Enabling} {Transparent} and {Reflective} {Decision}-{Making}}. In \bibinfo{booktitle}{\emph{Mensch und {Computer} 2023}}. \bibinfo{publisher}{ACM}, \bibinfo{address}{Rapperswil Switzerland}, \bibinfo{pages}{220--236}.
\newblock
\href{https://doi.org/10.1145/3603555.3603567}{doi:\nolinkurl{10.1145/3603555.3603567}}


\bibitem[Leiner(2024)]%
        {leiner2024soscisurvey}
\bibfield{author}{\bibinfo{person}{Dominik~J. Leiner}.} \bibinfo{year}{2024}\natexlab{}.
\newblock \bibinfo{booktitle}{\emph{SoSci Survey (Version 3.5.01)}}.
\newblock
\urldef\tempurl%
\url{https://www.soscisurvey.de}
\showURL{%
\tempurl}
\newblock
\shownote{Computer software}.


\bibitem[Lin and Lu(2011)]%
        {lin_why_2011}
\bibfield{author}{\bibinfo{person}{Kuan~Yu Lin} {and} \bibinfo{person}{Hsi~Peng Lu}.} \bibinfo{year}{2011}\natexlab{}.
\newblock \showarticletitle{Why people use social networking sites: An empirical study integrating network externalities and motivation theory}.
\newblock  \bibinfo{volume}{27}, \bibinfo{number}{3} (\bibinfo{year}{2011}), \bibinfo{pages}{1152--1161}.
\newblock
\href{https://doi.org/10.1016/j.chb.2010.12.009}{doi:\nolinkurl{10.1016/j.chb.2010.12.009}}
\newblock
\shownote{Publisher: Elsevier Ltd}.


\bibitem[Luepsen(2017)]%
        {luepsen_aligned_2017}
\bibfield{author}{\bibinfo{person}{Haiko Luepsen}.} \bibinfo{year}{2017}\natexlab{}.
\newblock \showarticletitle{The aligned rank transform and discrete variables: A warning}.
\newblock  \bibinfo{volume}{46}, \bibinfo{number}{9} (\bibinfo{year}{2017}), \bibinfo{pages}{6923--6936}.
\newblock
\showISSN{0361-0918, 1532-4141}
\href{https://doi.org/10.1080/03610918.2016.1217014}{doi:\nolinkurl{10.1080/03610918.2016.1217014}}


\bibitem[Lukoff et~al\mbox{.}(2021)]%
        {lukoff_how_2021}
\bibfield{author}{\bibinfo{person}{Kai Lukoff}, \bibinfo{person}{J~Vera Liao}, \bibinfo{person}{James Choi}, \bibinfo{person}{Kaiyue Fan}, \bibinfo{person}{Sean~A Munson}, {and} \bibinfo{person}{Alexis Hiniker}.} \bibinfo{year}{2021}\natexlab{}.
\newblock \showarticletitle{How the {Design} of {YouTube} {Influences} {User} {Sense} of {Agency}}. In \bibinfo{booktitle}{\emph{{CHI}'21}}. \bibinfo{publisher}{ACM}, \bibinfo{address}{Yokohama}.
\newblock
\showISBNx{978-1-4503-8096-6}
\href{https://doi.org/10.1145/3411764.3445467}{doi:\nolinkurl{10.1145/3411764.3445467}}


\bibitem[Mahdi et~al\mbox{.}(2017)]%
        {mahdi2017international}
\bibfield{author}{\bibinfo{person}{Soheil Mahdi}, \bibinfo{person}{Marisa Viljoen}, \bibinfo{person}{Rafael Massuti}, \bibinfo{person}{Melissa Selb}, \bibinfo{person}{Omar Almodayfer}, \bibinfo{person}{Sunil Karande}, \bibinfo{person}{Petrus~J de Vries}, \bibinfo{person}{Luis Rohde}, {and} \bibinfo{person}{Sven B{\"o}lte}.} \bibinfo{year}{2017}\natexlab{}.
\newblock \showarticletitle{An international qualitative study of ability and disability in ADHD using the WHO-ICF framework}.
\newblock \bibinfo{journal}{\emph{European child \& adolescent psychiatry}}  \bibinfo{volume}{26} (\bibinfo{year}{2017}), \bibinfo{pages}{1219--1231}.
\newblock


\bibitem[Maier and Harr(2020)]%
        {maier_dark_2020}
\bibfield{author}{\bibinfo{person}{Maximilian Maier} {and} \bibinfo{person}{Rikard Harr}.} \bibinfo{year}{2020}\natexlab{}.
\newblock \showarticletitle{Dark design patterns: An end-user perspective}.
\newblock  \bibinfo{volume}{16}, \bibinfo{number}{2} (\bibinfo{year}{2020}), \bibinfo{pages}{170--199}.
\newblock
\urldef\tempurl%
\url{https://ht.csr-pub.eu/index.php/ht/article/view/6}
\showURL{%
\tempurl}


\bibitem[Masi et~al\mbox{.}(2021)]%
        {masi2021video}
\bibfield{author}{\bibinfo{person}{Laura Masi}, \bibinfo{person}{Pascale Abadie}, \bibinfo{person}{Catherine Herba}, \bibinfo{person}{Mutsuko Emond}, \bibinfo{person}{Marie-Pier Gingras}, {and} \bibinfo{person}{Leila~Ben Amor}.} \bibinfo{year}{2021}\natexlab{}.
\newblock \showarticletitle{Video games in ADHD and non-ADHD children: Modalities of use and association with ADHD symptoms}.
\newblock \bibinfo{journal}{\emph{Frontiers in pediatrics}}  \bibinfo{volume}{9} (\bibinfo{year}{2021}), \bibinfo{pages}{632272}.
\newblock
\href{https://doi.org/10.3389/fped.2021.632272}{doi:\nolinkurl{10.3389/fped.2021.632272}}


\bibitem[Mathur et~al\mbox{.}(2019)]%
        {mathur_dark_2019}
\bibfield{author}{\bibinfo{person}{Arunesh Mathur}, \bibinfo{person}{Gunes Acar}, \bibinfo{person}{Michael~J. Friedman}, \bibinfo{person}{Elena Lucherini}, \bibinfo{person}{Jonathan Mayer}, \bibinfo{person}{Marshini Chetty}, {and} \bibinfo{person}{Arvind Narayanan}.} \bibinfo{year}{2019}\natexlab{}.
\newblock \showarticletitle{Dark {Patterns} at {Scale}: {Findings} from a {Crawl} of {11K} {Shopping} {Websites}}.
\newblock \bibinfo{journal}{\emph{Proceedings of the ACM on Human-Computer Interaction}} \bibinfo{volume}{3}, \bibinfo{number}{CSCW} (\bibinfo{date}{Nov.} \bibinfo{year}{2019}), \bibinfo{pages}{1--32}.
\newblock
\showISSN{2573-0142}
\href{https://doi.org/10.1145/3359183}{doi:\nolinkurl{10.1145/3359183}}


\bibitem[Mathur et~al\mbox{.}(2021)]%
        {mathur_what_2021}
\bibfield{author}{\bibinfo{person}{Arunesh Mathur}, \bibinfo{person}{Jonathan Mayer}, {and} \bibinfo{person}{Mihir Kshirsagar}.} \bibinfo{year}{2021}\natexlab{}.
\newblock \showarticletitle{What {Makes} a {Dark} {Pattern} ... {Dark} ? {Design} {Attributes}, {Normative} {Considerations}, and {Measurement} {Methods}}. In \bibinfo{booktitle}{\emph{{CHI}'21}}. \bibinfo{publisher}{ACM, New York, NY, USA}, \bibinfo{address}{Yokohama}, \bibinfo{pages}{18}.
\newblock
\showISBNx{978-1-4503-8096-6}
\href{https://doi.org/10.1145/3411764.3445610}{doi:\nolinkurl{10.1145/3411764.3445610}}


\bibitem[{McLennan}(2016)]%
        {mclennan_understanding_2016}
\bibfield{author}{\bibinfo{person}{John~D. {McLennan}}.} \bibinfo{year}{2016}\natexlab{}.
\newblock \showarticletitle{Understanding attention deficit hyperactivity disorder as a continuum}.
\newblock  \bibinfo{volume}{62}, \bibinfo{number}{12} (\bibinfo{year}{2016}), \bibinfo{pages}{979--982}.
\newblock
\showISSN{0008-350X}
\urldef\tempurl%
\url{https://www.ncbi.nlm.nih.gov/pmc/articles/PMC5154646/}
\showURL{%
\tempurl}


\bibitem[Mildner et~al\mbox{.}(2024a)]%
        {mildner_listening_2024}
\bibfield{author}{\bibinfo{person}{Thomas Mildner}, \bibinfo{person}{Orla Cooney}, \bibinfo{person}{Anna-Maria Meck}, \bibinfo{person}{Marion Bartl}, \bibinfo{person}{Gian-Luca Savino}, \bibinfo{person}{Philip~R Doyle}, \bibinfo{person}{Diego Garaialde}, \bibinfo{person}{Leigh Clark}, \bibinfo{person}{John Sloan}, \bibinfo{person}{Nina Wenig}, \bibinfo{person}{Rainer Malaka}, {and} \bibinfo{person}{Jasmin Niess}.} \bibinfo{year}{2024}\natexlab{a}.
\newblock \showarticletitle{Listening to the Voices: Describing Ethical Caveats of Conversational User Interfaces According to Experts and Frequent Users}. In \bibinfo{booktitle}{\emph{Proceedings of the CHI Conference on Human Factors in Computing Systems}} (Honolulu, HI, USA) \emph{(\bibinfo{series}{CHI '24})}. \bibinfo{publisher}{Association for Computing Machinery}, \bibinfo{address}{New York, NY, USA}, Article \bibinfo{articleno}{307}, \bibinfo{numpages}{18}~pages.
\newblock
\showISBNx{9798400703300}
\href{https://doi.org/10.1145/3613904.3642542}{doi:\nolinkurl{10.1145/3613904.3642542}}


\bibitem[Mildner et~al\mbox{.}(2025)]%
        {mildner_thomasmildnersns-research_2025}
\bibfield{author}{\bibinfo{person}{Thomas Mildner}, \bibinfo{person}{Daniel Fidel}, {and} \bibinfo{person}{Evropi Stefanidi}.} \bibinfo{year}{2025}\natexlab{}.
\newblock \bibinfo{title}{{ThomasMildner}/{SNS}-research: v1.0.0}.
\newblock
\href{https://doi.org/10.5281/zenodo.14761223}{doi:\nolinkurl{10.5281/zenodo.14761223}}
\newblock
\shownote{https://github.com/ThomasMildner/SNS-research/tree/CHI-2025}.


\bibitem[Mildner et~al\mbox{.}(2023a)]%
        {mildner_defending_2023}
\bibfield{author}{\bibinfo{person}{Thomas Mildner}, \bibinfo{person}{Merle Freye}, \bibinfo{person}{Gian-Luca Savino}, \bibinfo{person}{Philip~R. Doyle}, \bibinfo{person}{Benjamin~R. Cowan}, {and} \bibinfo{person}{Rainer Malaka}.} \bibinfo{year}{2023}\natexlab{a}.
\newblock \showarticletitle{{Defending Against the Dark Arts: Recognising Dark Patterns in Social Media}}. In \bibinfo{booktitle}{\emph{Designing Interactive Systems Conference (DIS '23), July 10--14, 2023, Pittsburgh, PA, USA}} (Pittsburgh, PA, USA) \emph{(\bibinfo{series}{DIS '23})}. \bibinfo{publisher}{Association for Computing Machinery}, \bibinfo{address}{New York, NY, USA}, \bibinfo{numpages}{13}~pages.
\newblock
\href{https://doi.org/1010.1145/3563657.3595964}{doi:\nolinkurl{1010.1145/3563657.3595964}}


\bibitem[Mildner and Savino(2021)]%
        {mildner_ethical_2021}
\bibfield{author}{\bibinfo{person}{Thomas Mildner} {and} \bibinfo{person}{Gian-Luca Savino}.} \bibinfo{year}{2021}\natexlab{}.
\newblock \showarticletitle{{Ethical User Interfaces: Exploring the Effects of Dark Patterns on Facebook}}. In \bibinfo{booktitle}{\emph{Extended Abstracts of the 2021 CHI Conference on Human Factors in Computing Systems}} (Yokohama, Japan) \emph{(\bibinfo{series}{CHI EA '21})}. \bibinfo{publisher}{Association for Computing Machinery}, \bibinfo{address}{New York, NY, USA}, Article \bibinfo{articleno}{464}, \bibinfo{numpages}{7}~pages.
\newblock
\showISBNx{978-1-4503-8095-9}
\href{https://doi.org/10.1145/3411763.3451659}{doi:\nolinkurl{10.1145/3411763.3451659}}


\bibitem[Mildner et~al\mbox{.}(2023b)]%
        {mildner_about_2023}
\bibfield{author}{\bibinfo{person}{Thomas Mildner}, \bibinfo{person}{Gian-Luca Savino}, \bibinfo{person}{Philip~R. Doyle}, \bibinfo{person}{Benjamin~R. Cowan}, {and} \bibinfo{person}{Rainer Malaka}.} \bibinfo{year}{2023}\natexlab{b}.
\newblock \showarticletitle{{About Engaging and Governing Strategies: A Thematic Analysis of Dark Patterns in Social Networking Services}}. In \bibinfo{booktitle}{\emph{Proceedings of the 2023 CHI Conference on Human Factors in Computing Systems}} (Hamburg, Germany) \emph{(\bibinfo{series}{CHI '23})}. \bibinfo{publisher}{Association for Computing Machinery}, \bibinfo{address}{New York, NY, USA}, Article \bibinfo{articleno}{192}, \bibinfo{numpages}{15}~pages.
\newblock
\showISBNx{9781450394215}
\href{https://doi.org/10.1145/3544548.3580695}{doi:\nolinkurl{10.1145/3544548.3580695}}


\bibitem[Mildner et~al\mbox{.}(2024b)]%
        {mildner_finding_2024}
\bibfield{author}{\bibinfo{person}{Thomas Mildner}, \bibinfo{person}{Gian-Luca Savino}, \bibinfo{person}{Susanne Putze}, {and} \bibinfo{person}{Rainer Malaka}.} \bibinfo{year}{2024}\natexlab{b}.
\newblock \showarticletitle{Finding a {Way} {Through} the {Social} {Media} {Labyrinth}: {Guiding} {Design} {Through} {User} {Expectations}}. In \bibinfo{booktitle}{\emph{Proceedings of the {International} {Conference} on {Mobile} and {Ubiquitous} {Multimedia}}} \emph{(\bibinfo{series}{{MUM} '24})}. \bibinfo{publisher}{Association for Computing Machinery}, \bibinfo{address}{New York, NY, USA}, \bibinfo{pages}{157--171}.
\newblock
\showISBNx{979-8-4007-1283-8}
\href{https://doi.org/10.1145/3701571.3701605}{doi:\nolinkurl{10.1145/3701571.3701605}}


\bibitem[Mildner et~al\mbox{.}(2024c)]%
        {mildner_dark_2024}
\bibfield{author}{\bibinfo{person}{Thomas Mildner}, \bibinfo{person}{Gian-Luca Savino}, \bibinfo{person}{Johannes Schöning}, {and} \bibinfo{person}{Rainer Malaka}.} \bibinfo{year}{2024}\natexlab{c}.
\newblock \showarticletitle{Dark {Patterns}: manipulative {Designstrategien} in digitalen {Gesundheitsanwendungen}}.
\newblock \bibinfo{journal}{\emph{Bundesgesundheitsblatt - Gesundheitsforschung - Gesundheitsschutz}}.
\newblock
\showISSN{1437-1588}
\href{https://doi.org/10.1007/s00103-024-03840-6}{doi:\nolinkurl{10.1007/s00103-024-03840-6}}


\bibitem[Monge~Roffarello et~al\mbox{.}(2023)]%
        {monge_roffarello_defining_2023}
\bibfield{author}{\bibinfo{person}{Alberto Monge~Roffarello}, \bibinfo{person}{Kai Lukoff}, {and} \bibinfo{person}{Luigi De~Russis}.} \bibinfo{year}{2023}\natexlab{}.
\newblock \showarticletitle{Defining and {Identifying} {Attention} {Capture} {Deceptive} {Designs} in {Digital} {Interfaces}}. In \bibinfo{booktitle}{\emph{Proceedings of the 2023 {CHI} {Conference} on {Human} {Factors} in {Computing} {Systems}}}. \bibinfo{publisher}{ACM}, \bibinfo{address}{Hamburg Germany}, \bibinfo{pages}{1--19}.
\newblock
\showISBNx{978-1-4503-9421-5}
\href{https://doi.org/10.1145/3544548.3580729}{doi:\nolinkurl{10.1145/3544548.3580729}}


\bibitem[Obi et~al\mbox{.}(2022)]%
        {obi_lets_2022}
\bibfield{author}{\bibinfo{person}{Ikechukwu Obi}, \bibinfo{person}{Colin~M. Gray}, \bibinfo{person}{Shruthi~Sai Chivukula}, \bibinfo{person}{Ja-Nae Duane}, \bibinfo{person}{Janna Johns}, \bibinfo{person}{Matthew Will}, \bibinfo{person}{Ziqing Li}, {and} \bibinfo{person}{Thomas Carlock}.} \bibinfo{year}{2022}\natexlab{}.
\newblock \showarticletitle{Let's {Talk} {About} {Socio}-{Technical} {Angst}: {Tracing} the {History} and {Evolution} of {Dark} {Patterns} on {Twitter} from 2010-2021}. \bibinfo{publisher}{arXiv}.
\newblock
\urldef\tempurl%
\url{http://arxiv.org/abs/2207.10563}
\showURL{%
\tempurl}


\bibitem[OECD(2022)]%
        {oecd__2022_dark}
\bibfield{author}{\bibinfo{person}{OECD}.} \bibinfo{year}{2022}\natexlab{}.
\newblock \bibinfo{title}{Dark commercial patterns}.
\newblock \bibinfo{numpages}{96}~pages.
\newblock


\bibitem[Organization(2018)]%
        {adhd_ICD}
\bibfield{author}{\bibinfo{person}{World~Health Organization}.} \bibinfo{year}{2018}\natexlab{}.
\newblock \bibinfo{booktitle}{\emph{ICD-11: International Classification of Diseases 11th Revision}}.
\newblock
\urldef\tempurl%
\url{https://icd.who.int/en}
\showURL{%
Retrieved July 4, 2021 from \tempurl}


\bibitem[Owens et~al\mbox{.}(2022)]%
        {owens_exploring_2022}
\bibfield{author}{\bibinfo{person}{Kentrell Owens}, \bibinfo{person}{Johanna Gunawan}, \bibinfo{person}{David Choffnes}, \bibinfo{person}{Pardis Emami-Naeini}, \bibinfo{person}{Tadayoshi Kohno}, {and} \bibinfo{person}{Franziska Roesner}.} \bibinfo{year}{2022}\natexlab{}.
\newblock \showarticletitle{Exploring Deceptive Design Patterns in Voice Interfaces}. In \bibinfo{booktitle}{\emph{Proceedings of the 2022 European Symposium on Usable Security}} (Karlsruhe Germany, 2022-09-29). \bibinfo{publisher}{{ACM}}, \bibinfo{pages}{64--78}.
\newblock
\showISBNx{978-1-4503-9700-1}
\href{https://doi.org/10.1145/3549015.3554213}{doi:\nolinkurl{10.1145/3549015.3554213}}


\bibitem[Pina et~al\mbox{.}(2014)]%
        {pina2014situ}
\bibfield{author}{\bibinfo{person}{Laura Pina}, \bibinfo{person}{Kael Rowan}, \bibinfo{person}{Asta Roseway}, \bibinfo{person}{Paul Johns}, \bibinfo{person}{Gillian~R. Hayes}, {and} \bibinfo{person}{Mary Czerwinski}.} \bibinfo{year}{2014}\natexlab{}.
\newblock \showarticletitle{In situ cues for ADHD parenting strategies using mobile technology}. In \bibinfo{booktitle}{\emph{Proceedings of the 8th International Conference on Pervasive Computing Technologies for Healthcare}} (Oldenburg, Germany) \emph{(\bibinfo{series}{PervasiveHealth '14})}. \bibinfo{publisher}{ICST (Institute for Computer Sciences, Social-Informatics and Telecommunications Engineering)}, \bibinfo{address}{Brussels, BEL}, \bibinfo{pages}{17–24}.
\newblock
\showISBNx{9781631900112}
\href{https://doi.org/10.4108/icst.pervasivehealth.2014.254958}{doi:\nolinkurl{10.4108/icst.pervasivehealth.2014.254958}}


\bibitem[Polanczyk et~al\mbox{.}(2007)]%
        {polanczyk2007worldwide}
\bibfield{author}{\bibinfo{person}{Guilherme Polanczyk}, \bibinfo{person}{Maur{\'\i}cio~Silva De~Lima}, \bibinfo{person}{Bernardo~Lessa Horta}, \bibinfo{person}{Joseph Biederman}, {and} \bibinfo{person}{Luis~Augusto Rohde}.} \bibinfo{year}{2007}\natexlab{}.
\newblock \showarticletitle{The worldwide prevalence of ADHD: a systematic review and metaregression analysis}.
\newblock \bibinfo{journal}{\emph{American journal of psychiatry}} \bibinfo{volume}{164}, \bibinfo{number}{6} (\bibinfo{year}{2007}), \bibinfo{pages}{942--948}.
\newblock


\bibitem[Prolific(2024)]%
        {prolific}
\bibfield{author}{\bibinfo{person}{Prolific}.} \bibinfo{year}{2024}\natexlab{}.
\newblock \bibinfo{booktitle}{\emph{Prolific {\textbar} Quickly find research participants you can trust}}.
\newblock
\urldef\tempurl%
\url{https://www.prolific.com}
\showURL{%
\tempurl}
\newblock
\shownote{Visited on 09-06-2024}.


\bibitem[Renaud et~al\mbox{.}(2024)]%
        {renaud2024we}
\bibfield{author}{\bibinfo{person}{Karen Renaud}, \bibinfo{person}{Cigdem Sengul}, \bibinfo{person}{Kovila Coopamootoo}, \bibinfo{person}{Bryan Clift}, \bibinfo{person}{Jacqui Taylor}, \bibinfo{person}{Mark Springett}, {and} \bibinfo{person}{Ben Morrison}.} \bibinfo{year}{2024}\natexlab{}.
\newblock \showarticletitle{“We’re Not That Gullible!” Revealing Dark Pattern Mental Models of 11-12-Year-Old Scottish Children}.
\newblock \bibinfo{journal}{\emph{ACM Transactions on Computer-Human Interaction}} \bibinfo{volume}{31}, \bibinfo{number}{3}, Article \bibinfo{articleno}{33} (\bibinfo{date}{aug} \bibinfo{year}{2024}), \bibinfo{numpages}{41}~pages.
\newblock
\showISSN{1073-0516}
\href{https://doi.org/10.1145/3660342}{doi:\nolinkurl{10.1145/3660342}}


\bibitem[Ruth et~al\mbox{.}(2013)]%
        {peroumental2013}
\bibfield{author}{\bibinfo{person}{Perou Ruth}, \bibinfo{person}{Bitsko~H Rebecca}, \bibinfo{person}{Blumberg~J Stephen}, \bibinfo{person}{Pastor Patricia}, \bibinfo{person}{Ghandour~M Reem}, \bibinfo{person}{Gfroerer~C Joseph}, \bibinfo{person}{Hedden~L Sarra}, \bibinfo{person}{Crosby~E Alex}, \bibinfo{person}{Visser~N Susanna}, \bibinfo{person}{Schieve~A Laura}, \bibinfo{person}{Parks~E Sharyn}, \bibinfo{person}{Hall~E Jeffery}, \bibinfo{person}{Brody Debra}, \bibinfo{person}{Simile~M Catherine}, \bibinfo{person}{Thompson~W William}, \bibinfo{person}{Baio Jon}, \bibinfo{person}{Avenevoli Shelli}, \bibinfo{person}{Kogan~D Michael}, \bibinfo{person}{Huang~N Larke}, \bibinfo{person}{Centers for Disease~Control}, {and} \bibinfo{person}{Prevention (CDC)}.} \bibinfo{year}{2013}\natexlab{}.
\newblock \showarticletitle{Mental health surveillance among children--United States, 2005-2011}.
\newblock \bibinfo{journal}{\emph{MMWR Surveill Summ 62 Suppl 2: 1–35}} (\bibinfo{year}{2013}).
\newblock


\bibitem[Sanchez~Chamorro et~al\mbox{.}(2024)]%
        {sanchez_chamorro_my_2024}
\bibfield{author}{\bibinfo{person}{Lorena Sanchez~Chamorro}, \bibinfo{person}{Carine Lallemand}, {and} \bibinfo{person}{Colin~M. Gray}.} \bibinfo{year}{2024}\natexlab{}.
\newblock \showarticletitle{"My Mother Told Me These Things are Always Fake" - Understanding Teenagers' Experiences with Manipulative Designs}. In \bibinfo{booktitle}{\emph{Designing Interactive Systems Conference}} ({IT} University of Copenhagen Denmark, 2024-07). \bibinfo{publisher}{{ACM}}, \bibinfo{pages}{1469--1482}.
\newblock
\showISBNx{9798400705830}
\href{https://doi.org/10.1145/3643834.3660704}{doi:\nolinkurl{10.1145/3643834.3660704}}


\bibitem[Schaffner et~al\mbox{.}(2022)]%
        {schaffner_understanding_2022}
\bibfield{author}{\bibinfo{person}{Brennan Schaffner}, \bibinfo{person}{Neha~A. Lingareddy}, {and} \bibinfo{person}{Marshini Chetty}.} \bibinfo{year}{2022}\natexlab{}.
\newblock \showarticletitle{Understanding {Account} {Deletion} and {Relevant} {Dark} {Patterns} on {Social} {Media}}.
\newblock \bibinfo{journal}{\emph{Proceedings of the ACM on Human-Computer Interaction}} \bibinfo{volume}{6}, \bibinfo{number}{CSCW2} (\bibinfo{date}{Nov.} \bibinfo{year}{2022}), \bibinfo{pages}{1--43}.
\newblock
\showISSN{2573-0142}
\href{https://doi.org/10.1145/3555142}{doi:\nolinkurl{10.1145/3555142}}


\bibitem[Schippers et~al\mbox{.}(2024)]%
        {schippers2024associations}
\bibfield{author}{\bibinfo{person}{Lessa~M Schippers}, \bibinfo{person}{CU Greven}, {and} \bibinfo{person}{M Hoogman}.} \bibinfo{year}{2024}\natexlab{}.
\newblock \showarticletitle{Associations between ADHD traits and self-reported strengths in the general population}.
\newblock \bibinfo{journal}{\emph{Comprehensive Psychiatry}}  \bibinfo{volume}{130} (\bibinfo{year}{2024}), \bibinfo{pages}{152461}.
\newblock


\bibitem[Sedgwick et~al\mbox{.}(2019)]%
        {sedgwick2019positive}
\bibfield{author}{\bibinfo{person}{Jane~Ann Sedgwick}, \bibinfo{person}{Andrew Merwood}, {and} \bibinfo{person}{Philip Asherson}.} \bibinfo{year}{2019}\natexlab{}.
\newblock \showarticletitle{The positive aspects of attention deficit hyperactivity disorder: a qualitative investigation of successful adults with ADHD}.
\newblock \bibinfo{journal}{\emph{ADHD Attention Deficit and Hyperactivity Disorders}} \bibinfo{volume}{11}, \bibinfo{number}{3} (\bibinfo{year}{2019}), \bibinfo{pages}{241--253}.
\newblock


\bibitem[Silva et~al\mbox{.}(2023)]%
        {silva2023unpacking}
\bibfield{author}{\bibinfo{person}{Lucas~M Silva}, \bibinfo{person}{Franceli~L Cibrian}, \bibinfo{person}{Elissa Monteiro}, \bibinfo{person}{Arpita Bhattacharya}, \bibinfo{person}{Jesus~A Beltran}, \bibinfo{person}{Clarisse Bonang}, \bibinfo{person}{Daniel~A Epstein}, \bibinfo{person}{Sabrina~EB Schuck}, \bibinfo{person}{Kimberley~D Lakes}, {and} \bibinfo{person}{Gillian~R Hayes}.} \bibinfo{year}{2023}\natexlab{}.
\newblock \showarticletitle{Unpacking the Lived Experiences of Smartwatch Mediated Self and Co-Regulation with ADHD Children}. In \bibinfo{booktitle}{\emph{Proceedings of the 2023 CHI Conference on Human Factors in Computing Systems}}. \bibinfo{pages}{1--19}.
\newblock


\bibitem[Sina et~al\mbox{.}(2022)]%
        {sina_social_2022}
\bibfield{author}{\bibinfo{person}{Elida Sina}, \bibinfo{person}{Daniel Boakye}, \bibinfo{person}{Lara Christianson}, \bibinfo{person}{Wolfgang Ahrens}, {and} \bibinfo{person}{Antje Hebestreit}.} \bibinfo{year}{2022}\natexlab{}.
\newblock \showarticletitle{Social Media and Children's and Adolescents' Diets: A Systematic Review of the Underlying Social and Physiological Mechanisms}.
\newblock  \bibinfo{volume}{13}, \bibinfo{number}{3} (\bibinfo{year}{2022}), \bibinfo{pages}{913--937}.
\newblock
\showISSN{2161-8313}
\href{https://doi.org/10.1093/advances/nmac018}{doi:\nolinkurl{10.1093/advances/nmac018}}


\bibitem[Sinclair and Grieve(2017)]%
        {sinclair2017facebook}
\bibfield{author}{\bibinfo{person}{Tara~J Sinclair} {and} \bibinfo{person}{Rachel Grieve}.} \bibinfo{year}{2017}\natexlab{}.
\newblock \showarticletitle{Facebook as a source of social connectedness in older adults}.
\newblock \bibinfo{journal}{\emph{Computers in Human Behavior}}  \bibinfo{volume}{66} (\bibinfo{year}{2017}), \bibinfo{pages}{363--369}.
\newblock


\bibitem[Sinders(2022)]%
        {sinders_whats_2022}
\bibfield{author}{\bibinfo{person}{Caroline Sinders}.} \bibinfo{year}{2022}\natexlab{}.
\newblock \bibinfo{booktitle}{\emph{What’s {In} a {Name}?}}
\newblock
\urldef\tempurl%
\url{https://medium.com/@carolinesinders/whats-in-a-name-unpacking-dark-patterns-versus-deceptive-design-e96068627ec4}
\showURL{%
\tempurl}
\newblock
\shownote{Accessed: 09/02/2024}.


\bibitem[Sonne et~al\mbox{.}(2016a)]%
        {sonne2016assistive}
\bibfield{author}{\bibinfo{person}{Tobias Sonne}, \bibinfo{person}{Paul Marshall}, \bibinfo{person}{Carsten Obel}, \bibinfo{person}{Per~Hove Thomsen}, {and} \bibinfo{person}{Kaj Gr\o{}nb\ae{}k}.} \bibinfo{year}{2016}\natexlab{a}.
\newblock \showarticletitle{An assistive technology design framework for ADHD}. In \bibinfo{booktitle}{\emph{Proceedings of the 28th Australian Conference on Computer-Human Interaction}} (Launceston, Tasmania, Australia) \emph{(\bibinfo{series}{OzCHI '16})}. \bibinfo{publisher}{Association for Computing Machinery}, \bibinfo{address}{New York, NY, USA}, \bibinfo{pages}{60–70}.
\newblock
\showISBNx{9781450346184}
\href{https://doi.org/10.1145/3010915.3010925}{doi:\nolinkurl{10.1145/3010915.3010925}}


\bibitem[Sonne et~al\mbox{.}(2016b)]%
        {sonne2016changing}
\bibfield{author}{\bibinfo{person}{Tobias Sonne}, \bibinfo{person}{J\"{o}rg M\"{u}ller}, \bibinfo{person}{Paul Marshall}, \bibinfo{person}{Carsten Obel}, {and} \bibinfo{person}{Kaj Gr\o{}nb\ae{}k}.} \bibinfo{year}{2016}\natexlab{b}.
\newblock \showarticletitle{Changing Family Practices with Assistive Technology: MOBERO Improves Morning and Bedtime Routines for Children with ADHD} \emph{(\bibinfo{series}{CHI '16})}. \bibinfo{publisher}{Association for Computing Machinery}, \bibinfo{address}{New York, NY, USA}, \bibinfo{pages}{152–164}.
\newblock
\showISBNx{9781450333627}
\href{https://doi.org/10.1145/2858036.2858157}{doi:\nolinkurl{10.1145/2858036.2858157}}


\bibitem[Sonne et~al\mbox{.}(2015)]%
        {sonne2015designing}
\bibfield{author}{\bibinfo{person}{Tobias Sonne}, \bibinfo{person}{Carsten Obel}, {and} \bibinfo{person}{Kaj Gr{\o}nb{\ae}k}.} \bibinfo{year}{2015}\natexlab{}.
\newblock \showarticletitle{Designing real time assistive technologies: a study of children with ADHD}. In \bibinfo{booktitle}{\emph{Proceedings of the Annual Meeting of the Australian Special Interest Group for Computer Human Interaction}}. \bibinfo{pages}{34--38}.
\newblock


\bibitem[Spiel et~al\mbox{.}(2019)]%
        {spiel2019agency}
\bibfield{author}{\bibinfo{person}{Katta Spiel}, \bibinfo{person}{Christopher Frauenberger}, \bibinfo{person}{Os Keyes}, {and} \bibinfo{person}{Geraldine Fitzpatrick}.} \bibinfo{year}{2019}\natexlab{}.
\newblock \showarticletitle{Agency of autistic children in technology research—A critical literature review}.
\newblock \bibinfo{journal}{\emph{ACM Transactions on Computer-Human Interaction (TOCHI)}} \bibinfo{volume}{26}, \bibinfo{number}{6} (\bibinfo{year}{2019}), \bibinfo{pages}{1--40}.
\newblock


\bibitem[Spiel and Gerling(2021)]%
        {spiel2021purpose}
\bibfield{author}{\bibinfo{person}{Katta Spiel} {and} \bibinfo{person}{Kathrin Gerling}.} \bibinfo{year}{2021}\natexlab{}.
\newblock \showarticletitle{The Purpose of Play: How HCI Games Research Fails Neurodivergent Populations}.
\newblock \bibinfo{journal}{\emph{ACM Transactions on Computer-Human Interaction (TOCHI)}} \bibinfo{volume}{28}, \bibinfo{number}{2} (\bibinfo{year}{2021}), \bibinfo{pages}{1--40}.
\newblock


\bibitem[Spiel et~al\mbox{.}(2022)]%
        {spiel2022adhd}
\bibfield{author}{\bibinfo{person}{Katta Spiel}, \bibinfo{person}{Eva Hornecker}, \bibinfo{person}{Rua~Mae Williams}, {and} \bibinfo{person}{Judith Good}.} \bibinfo{year}{2022}\natexlab{}.
\newblock \showarticletitle{ADHD and technology research--investigated by neurodivergent readers}. In \bibinfo{booktitle}{\emph{Proceedings of the 2022 CHI Conference on Human Factors in Computing Systems}}. \bibinfo{pages}{1--21}.
\newblock


\bibitem[Stefanidi et~al\mbox{.}(2021)]%
        {stefanidi2021children}
\bibfield{author}{\bibinfo{person}{Evropi Stefanidi}, \bibinfo{person}{Maria Korozi}, \bibinfo{person}{Asterios Leonidis}, \bibinfo{person}{Dimitrios Arampatzis}, \bibinfo{person}{Margherita Antona}, {and} \bibinfo{person}{George Papagiannakis}.} \bibinfo{year}{2021}\natexlab{}.
\newblock \showarticletitle{When children program intelligent environments: Lessons learned from a serious AR game}. In \bibinfo{booktitle}{\emph{Interaction Design and Children}}. \bibinfo{pages}{375--386}.
\newblock


\bibitem[Stefanidi et~al\mbox{.}(2022)]%
        {stefanidi2022designing}
\bibfield{author}{\bibinfo{person}{Evropi Stefanidi}, \bibinfo{person}{Johannes Sch\"{o}ning}, \bibinfo{person}{Sebastian~S. Feger}, \bibinfo{person}{Paul Marshall}, \bibinfo{person}{Yvonne Rogers}, {and} \bibinfo{person}{Jasmin Niess}.} \bibinfo{year}{2022}\natexlab{}.
\newblock \showarticletitle{Designing for Care Ecosystems: a Literature Review of Technologies for Children with ADHD}. In \bibinfo{booktitle}{\emph{Proceedings of the 21st Annual ACM Interaction Design and Children Conference}} (Braga, Portugal) \emph{(\bibinfo{series}{IDC '22})}. \bibinfo{publisher}{Association for Computing Machinery}, \bibinfo{address}{New York, NY, USA}, \bibinfo{pages}{13–25}.
\newblock
\showISBNx{9781450391979}
\href{https://doi.org/10.1145/3501712.3529746}{doi:\nolinkurl{10.1145/3501712.3529746}}


\bibitem[Stefanidi et~al\mbox{.}(2023)]%
        {stefanidi2023children}
\bibfield{author}{\bibinfo{person}{Evropi Stefanidi}, \bibinfo{person}{Johannes Sch{\"o}ning}, \bibinfo{person}{Yvonne Rogers}, {and} \bibinfo{person}{Jasmin Niess}.} \bibinfo{year}{2023}\natexlab{}.
\newblock \showarticletitle{Children with ADHD and their Care Ecosystem: Designing Beyond Symptoms}. In \bibinfo{booktitle}{\emph{Proceedings of the 2023 CHI Conference on Human Factors in Computing Systems}}. \bibinfo{pages}{1--17}.
\newblock
\href{https://doi.org/10.1145/3544548.3581216}{doi:\nolinkurl{10.1145/3544548.3581216}}


\bibitem[Stefanidi et~al\mbox{.}(2024)]%
        {stefanidi2024moodgems}
\bibfield{author}{\bibinfo{person}{Evropi Stefanidi}, \bibinfo{person}{Jonathan~L.B Wassmann}, \bibinfo{person}{Pawe\l{}~W. Wo\'{z}niak}, \bibinfo{person}{Gunnar Spellmeyer}, \bibinfo{person}{Yvonne Rogers}, {and} \bibinfo{person}{Jasmin Niess}.} \bibinfo{year}{2024}\natexlab{}.
\newblock \showarticletitle{MoodGems: Designing for theWell-being of Children with ADHD and their Families at Home}. In \bibinfo{booktitle}{\emph{Proceedings of the 23rd Annual ACM Interaction Design and Children Conference}} \emph{(\bibinfo{series}{IDC '24})}. \bibinfo{publisher}{Association for Computing Machinery}, \bibinfo{address}{New York, NY, USA}, \bibinfo{numpages}{15}~pages.
\newblock
\href{https://doi.org/10.1145/3628516.3655795}{doi:\nolinkurl{10.1145/3628516.3655795}}


\bibitem[Wehmeier et~al\mbox{.}(2010)]%
        {wehmeier2010social}
\bibfield{author}{\bibinfo{person}{Peter~M Wehmeier}, \bibinfo{person}{Alexander Schacht}, {and} \bibinfo{person}{Russell~A Barkley}.} \bibinfo{year}{2010}\natexlab{}.
\newblock \showarticletitle{Social and emotional impairment in children and adolescents with ADHD and the impact on quality of life}.
\newblock \bibinfo{journal}{\emph{Journal of Adolescent health}} \bibinfo{volume}{46}, \bibinfo{number}{3} (\bibinfo{year}{2010}), \bibinfo{pages}{209--217}.
\newblock


\bibitem[Wender et~al\mbox{.}(2001)]%
        {wender2001adults}
\bibfield{author}{\bibinfo{person}{Paul~H Wender}, \bibinfo{person}{Lorraine~E Wolf}, {and} \bibinfo{person}{Jeanette Wasserstein}.} \bibinfo{year}{2001}\natexlab{}.
\newblock \showarticletitle{Adults with ADHD: An overview}.
\newblock \bibinfo{journal}{\emph{Annals of the New York academy of sciences}} \bibinfo{volume}{931}, \bibinfo{number}{1} (\bibinfo{year}{2001}), \bibinfo{pages}{1--16}.
\newblock


\bibitem[Werling et~al\mbox{.}(2022)]%
        {werling2022problematic}
\bibfield{author}{\bibinfo{person}{Anna~Maria Werling}, \bibinfo{person}{Sajiv Kuzhippallil}, \bibinfo{person}{Sophie Emery}, \bibinfo{person}{Susanne Walitza}, {and} \bibinfo{person}{Renate Drechsler}.} \bibinfo{year}{2022}\natexlab{}.
\newblock \showarticletitle{Problematic use of digital media in children and adolescents with a diagnosis of attention-deficit/hyperactivity disorder compared to controls. A meta-analysis}.
\newblock \bibinfo{journal}{\emph{Journal of Behavioral Addictions}} (\bibinfo{year}{2022}), \bibinfo{pages}{305--325}.
\newblock
\href{https://doi.org/10.1556/2006.2022.00007}{doi:\nolinkurl{10.1556/2006.2022.00007}}


\bibitem[White and Shah(2006)]%
        {white2006uninhibited}
\bibfield{author}{\bibinfo{person}{Holly~A White} {and} \bibinfo{person}{Priti Shah}.} \bibinfo{year}{2006}\natexlab{}.
\newblock \showarticletitle{Uninhibited imaginations: creativity in adults with attention-deficit/hyperactivity disorder}.
\newblock \bibinfo{journal}{\emph{Personality and individual differences}} \bibinfo{volume}{40}, \bibinfo{number}{6} (\bibinfo{year}{2006}), \bibinfo{pages}{1121--1131}.
\newblock


\bibitem[Wobbrock et~al\mbox{.}(2011)]%
        {wobbock_align_2011}
\bibfield{author}{\bibinfo{person}{Jacob~O. Wobbrock}, \bibinfo{person}{Leah Findlater}, \bibinfo{person}{Darren Gergle}, {and} \bibinfo{person}{James~J. Higgins}.} \bibinfo{year}{2011}\natexlab{}.
\newblock \showarticletitle{The aligned rank transform for nonparametric factorial analyses using only anova procedures}. In \bibinfo{booktitle}{\emph{Proceedings of the SIGCHI Conference on Human Factors in Computing Systems}} (Vancouver, BC, Canada) \emph{(\bibinfo{series}{CHI '11})}. \bibinfo{publisher}{Association for Computing Machinery}, \bibinfo{address}{New York, NY, USA}, \bibinfo{pages}{143–146}.
\newblock
\showISBNx{9781450302289}
\href{https://doi.org/10.1145/1978942.1978963}{doi:\nolinkurl{10.1145/1978942.1978963}}


\bibitem[Zagal et~al\mbox{.}(2013)]%
        {zagal_dark_2013}
\bibfield{author}{\bibinfo{person}{José~P Zagal}, \bibinfo{person}{Staffan Björk}, {and} \bibinfo{person}{Chris Lewis}.} \bibinfo{year}{2013}\natexlab{}.
\newblock \showarticletitle{Dark Patterns in the Design of Games}. In \bibinfo{booktitle}{\emph{Proceedings of the 8th International Conference on the Foundations of Digital Games (FDG 2013)}} (May 14-17). \bibinfo{publisher}{Society for the Advancement of the Science of Digital Games}, \bibinfo{address}{Chania, Crete, Greece}, \bibinfo{pages}{39--46}.
\newblock
\showISBNx{978-0-9913982-0-1}
\urldef\tempurl%
\url{http://www.fdg2013.org/program/papers.html}
\showURL{%
\tempurl}


\bibitem[Zasler et~al\mbox{.}(2013)]%
        {zasler2013neurobehavioral}
\bibfield{author}{\bibinfo{person}{Nathan~D Zasler}, \bibinfo{person}{Michael~F Martelli}, {and} \bibinfo{person}{Harvey~E Jacobs}.} \bibinfo{year}{2013}\natexlab{}.
\newblock \showarticletitle{Neurobehavioral disorders}.
\newblock \bibinfo{journal}{\emph{Handbook of clinical neurology}}  \bibinfo{volume}{110} (\bibinfo{year}{2013}), \bibinfo{pages}{377--388}.
\newblock


\bibitem[Zuboff(2023)]%
        {zuboff_surveillance_2023}
\bibfield{author}{\bibinfo{person}{Shoshana Zuboff}.} \bibinfo{year}{2023}\natexlab{}.
\newblock \showarticletitle{The age of surveillance capitalism}. In \bibinfo{booktitle}{\emph{Social Theory Re-Wired}}. \bibinfo{publisher}{Routledge}, \bibinfo{pages}{203--213}.
\newblock


\end{thebibliography}
